\documentclass[superscriptaddress,
twocolumn,
nofootinbib,
amsmath,amssymb,
aps,
pra,
floatfix,eqsecnum
]{revtex4-2}
\usepackage{amsmath,amsthm,amssymb,graphicx,braket,bm}
\usepackage[
colorlinks=true,
linkcolor=black,
citecolor=black,
urlcolor=black
]{hyperref}
\usepackage{ragged2e}
\usepackage{xcolor}
\usepackage{orcidlink}
\begin{document}

\title{Preprocessing noise in finite-size quantum key distribution}

\author{Gabriele Staffieri~\orcidlink{0009-0004-8312-593X}}
\affiliation{Dipartimento Interateneo di Fisica, Università di Bari, 70126 Bari, Italy}
\affiliation{INFN, Sezione di Bari, 70126 Bari, Italy}

\author{Giuseppe D'Ambruoso~\orcidlink{0009-0003-0517-2901}}
\affiliation{INFN, Sezione di Bari, 70126 Bari, Italy}
\affiliation{Dipartimento di Meccanica, Matematica e Management, Politecnico di Bari, 70125 Bari, Italy}
\affiliation{Dipartimento Interateneo di Fisica, Politecnico di Bari, 70126 Bari, Italy}

\author{Giovanni Scala~\orcidlink{0000-0003-2685-0946}}
\affiliation{INFN, Sezione di Bari, 70126 Bari, Italy}
\affiliation{Dipartimento Interateneo di Fisica, Politecnico di Bari, 70126 Bari, Italy}

\author{Cosmo Lupo~\orcidlink{0000-0002-5227-4009}}
\affiliation{Dipartimento Interateneo di Fisica, Università di Bari, 70126 Bari, Italy}
\affiliation{INFN, Sezione di Bari, 70126 Bari, Italy}
\affiliation{Dipartimento Interateneo di Fisica, Politecnico di Bari, 70126 Bari, Italy}

\begin{abstract}
It is known that trusted, preprocessing noise may boost quantum key distribution by expanding the range of values of tolerated noise. For BB84, adding trusted noise may allow the generation of secret keys even for qubit error rate (QBER) beyond the $11\%$ threshold in the asymptotic regime.
Here we study the effect of preprocessing noise in the finite-size regime where only a limited number of signals are exchanged between Alice and Bob.
{We compute the secret key rate from the full variational formulation of the sandwiched R\'enyi conditional entropy, optimized through nonsymmetric conic optimization.}
We find that trusted noise improves the key rate only for a finite interval of $\alpha$, from the $\alpha \to 1$ limit up to $\alpha \simeq 1.4 $.
{We then optimize the finite-size key rate jointly over both $q$, $\alpha$, and the fraction of raw keys used for parameter estimation, to study how trusted preprocessing changes the maximum tolerable QBER as the block size increases. We show that the secret-key rate enhancement provided by optimized noisy preprocessing becomes more pronounced at higher QBER levels and can increase the maximum tolerable QBER over a broad range of finite block sizes.}
\end{abstract}

\maketitle

\section{Introduction} 

Quantum key distribution (QKD) enables two distant users, commonly dubbed Alice and Bob, to establish a shared secret key, with information-theoretic security, based on the principles of quantum physics and suitable additional assumptions, even against adversaries with unlimited quantum computational power~\cite{Gisin2002,Pirandola2020,Portmann2022,Primaatmaja2023,Ghoreishi2025a}. Nowadays security analyses typically adopt composable frameworks that takes into account finite-size correction, using one-shot entropic quantities~\cite{renner2008security,tomamichel2015quantum}.
In discrete-variable (DV) protocols, information is encoding in discrete degrees of freedom of the quantum electromagnetic fields, and measurements are usually performed on mutually unbiased bases. Among the various DV protocols, the BB84 is the most well-known example~\cite{bennett2014quantum}, which has played a fundamental role for the development of experiments and security proofs~\cite{shor2000simple}. Composable security proofs offer general bounds on the finite-size key rates and clarify the impact of classical preprocessing and postprocessing~\cite{renner2005information,mertz2013quantum}. 
One important example is \emph{trusted local randomization}, where one party flips each bit of the sifted key with a given probability before information reconciliation. This preprocessing is often referred to as \emph{trusted-noise injection}; in this way, one of the legitimate user adds local randomness which is inaccessible to a possible eavesdropper. 
In the asymptotic limit where an unbounded number of signals is exchanged, this technique has been proven to increase the secret-key rate, especially when the quantum bit error rate (QBER) is high, as shown in early studies~\cite{kraus2005lower,renes2007noisy}.

Recent work on finite-size security indicates that using {Rényi entropies} can lead to tighter key-rate bounds for practical block lengths~\cite{dupuis2023privacy}. In DV QKD, several frameworks based on Rényi entropy and entropy accumulation have been developed and applied—for example, in decoy-state protocols~\cite{Kamin_2025,kamin2025r}. Furthermore, similar techniques have also been explored in continuous-variable (CV) settings, including CV QKD with discrete modulation under Gaussian attacks~\cite{staffieri2026finiteCV,yamano2023general} or general attacks \cite{navarro2025finite,Navarro2026DMQKD}, while. More recently, R\'enyi-based finite-size security techniques
have also been applied to relativistic phase-shift-keying
(RPSK) QKD, combining variable-length entropy accumulation
with conic optimization to establish security against
general attacks~\cite{PascualGarcia2026RPSK}.
As a matter of fact, sandwiched Rényi entropies provide a direct and tight quantification of the secret-key generation rate without passing through the procedure of min-entropy smoothing~\cite{dupuis2023privacy,george2025finite}.

These developments motivate a deeper analysis on the role of trusted noise in the finite-size regime. Here we develop this analysis in terms of sandwiched R\'enyi entropies.
In contrast with other entropic quantities, such as the von Neumann entropy and the min-entropy, R\'enyi-based bounds respond differently to noise levels and introduce the R\'enyi order $\alpha$ as an additional optimization parameter.
This may affect the trade-off between the privacy gain from trusted noise and the increased information leakage from error reconciliation. As a consequence, the expected key-rate enhancement may no longer hold in all regimes.

In this work, we analyze the security of BB84, with trusted preprocessing noise, against collective (i.i.d.)~attacks, in the finite-size regime {(for a comparison with the six-state protocol see also Appendix \ref{app:sixstate})}.
First, we identify the range of values of the R\'enyi order $\alpha$ for which trusted noise remains beneficial, through a comparison of the key rates with and without optimally tuned randomization.
{Second, we determine finite-size secret-key rates by optimizing over both the trusted randomization parameter $q$ and the R\'enyi order $\alpha$.}
Third, we investigate the maximum tolerable values of the QBER as a function of the block size.
{The central numerical ingredient is the full variational R\'enyi optimization over both the bipartite state $\rho_{AB}$ and the auxiliary state $\sigma$. We evaluate this quantity through the nonsymmetric R\'enyiQKD conic formulation of Ref.~\cite{navarro2025finite,lorente2025quantum}, which is particularly well suited to the regime in which the R\'enyi entropy approaches the von Neumann limit.}
{Finally, as a benchmark, we develop a complete two-variable Frank--Wolfe formulation of the same variational objective, including analytical gradients with respect to both optimization variables. }

\section{Methodology}
\label{sec:method}

\subsection{Protocol model (BB84 with trusted local randomization)}

\begin{figure*}[t] 
    \centering
    \includegraphics[width=0.8\textwidth]{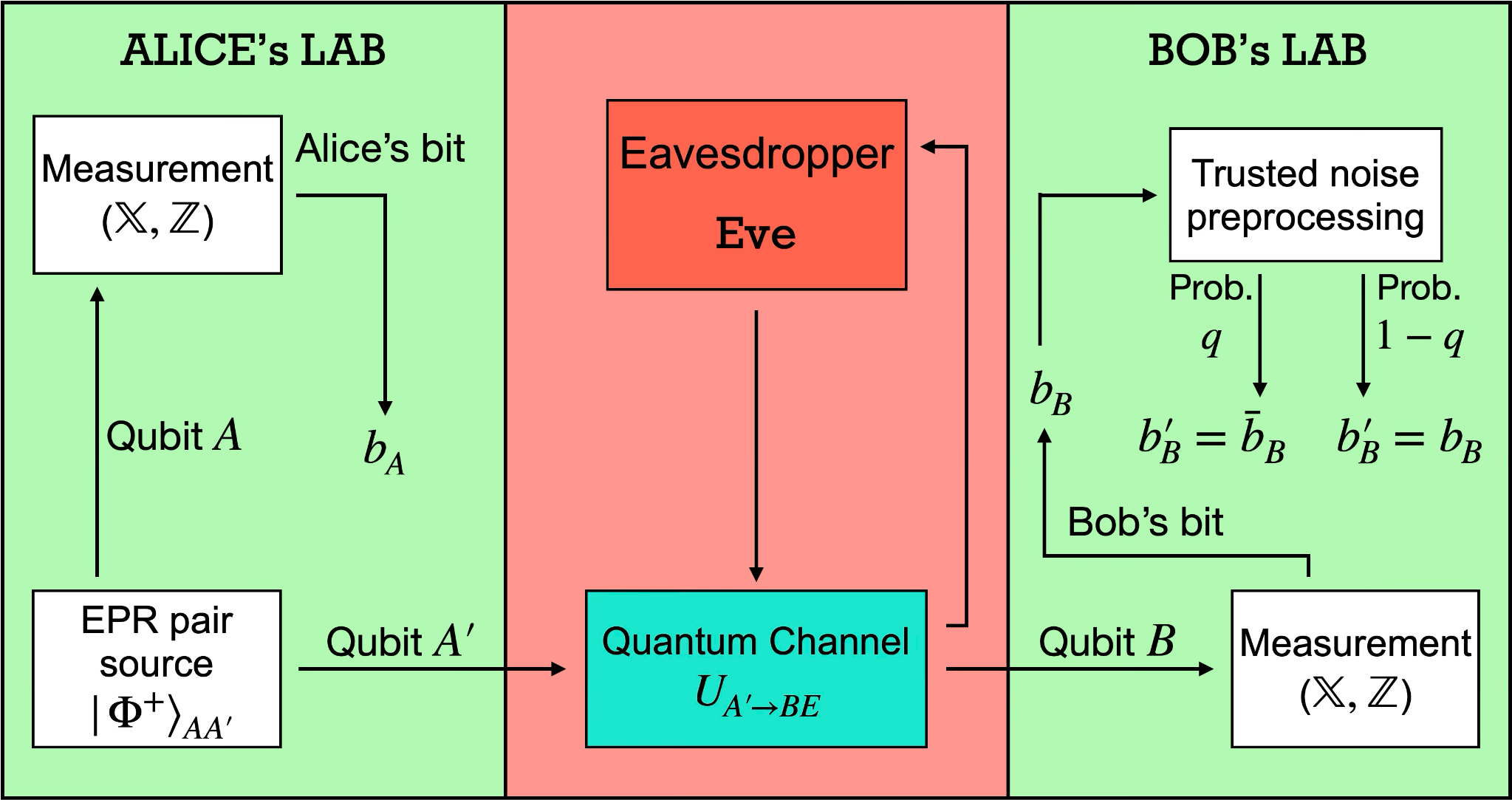}
    \caption{\justifying Schematic of BB84 protocol with trusted noise preprocessing, in the entanglement-based representation. Alice and Bob's classical bits obtained though qubit measurements are indicated as $b_A$ and $b_B$ respectively. The symbol $b'_B$ indicates Bob's bit after trusted bit flip.}
    \label{prot_scheme}
\end{figure*}

To assess the security of the protocol, we consider the entanglement-based (EB) representation of the BB84 protocol, with the introduction of trusted preprocessing noise as in Fig. \ref{prot_scheme}. 
Furthermore, we work under the assumption of collective attacks, where each quantum signal transmission from Alice to Bob is subject to i.i.d.~noise.
Before proceeding, we briefly recall a high-level description of the protocol:
\begin{enumerate}
    \item \textit{Preparation and transmission:} Alice prepares $n$ identical maximally entangled qubit pairs $|\Phi^+\rangle_{AA'} = (|0\rangle_A |0\rangle_{A'} + |1 \rangle_A |1 \rangle_{A'})/\sqrt{2}$. She keeps qubit $A$ and sends qubit $A'$ to Bob.
    \item{Measurement:} in each round, Alice and Bob measure in randomly chosen $\mathbb{Z} = \{ |0\rangle, |1\rangle\}$ or $\mathbb{X} = \{ |\texttt{+}\rangle, |\texttt{-}\rangle \}$ basis.
    \item \textit{Sifting:} they publicly announce the bases and keep only matched-basis rounds, obtaining {$N$ sifted bits before the test samples are revealed} out of $n$ transmissions.
    {\item \textit{Parameter estimation:} Alice and Bob reveal $k_X$ and $k_Z$ sifted bits to estimate the error rates in the two bases. The number of sifted rounds remaining for key generation is $m=N-k_X-k_Z$.}
    \item \textit{Trusted local randomization:} before information reconciliation, Bob flips each {remaining} sifted bit independently with probability $q$ (hence introducing trusted noise).
    \item \textit{Information reconciliation:} in reverse reconciliation Bob reveals {$\ell_{\rm leak}$} bits of syndrome information.
    \item {\textit{Error verification:} Alice and Bob compare a two-universal hash of their reconciled strings. A verification tag of length $t_{\rm EC}$ bounds the probability of accepting different keys by $\epsilon_{\rm EC}$.}
    \item \textit{Privacy amplification:} a two-universal hash function extracts an $\ell$-bit key from the reconciled strings.
\end{enumerate}

Under the assumption of collective attacks, each signal transmission is described by an isometry $U_{A' \to BE}$ mapping Alice qubit $A'$ into Bob qubit $B$ and the environment $E$. The latter is without loss of generality prepared in a given state $|\phi\rangle_E$. The tri-partite state shared by Alice, Bob, and the environment, after step 1.~of the protocol is
\begin{align}
|\Theta\rangle_{ABE} =
(\mathbb{I}_A\otimes U_{A' \to BE})\,
|\Phi^+\rangle_{AA'} |\phi\rangle_E \, .
\end{align}

The bipartite state shared by Alice and Bob after each transmission is 
$\rho_{AB} = \mathrm{tr}_E\!\left(|\Theta\rangle\langle\Theta|\right)$.
For collective attacks, the $n$-round state is the tensor power, 
$\rho_{A^nB^n}=\rho_{AB}^{\otimes n}$, where $A^n=A_1\cdots A_n$ and $B^n=B_1\cdots B_n$. 
The channel induced by $U_{A'\to BE}$ is determined by the adversary (denoted Eve) who controls and monitors the environment, and is unknown to the legitimate parties.

Let $Z$ denote the random variable representing Bob's classical bit produced by his measurement in the $\mathbb{Z}$ basis. 
Trusted noise is modeled as a local, random bit-flip,
\begin{align}
Z \ \mapsto\ Y_1 = Z\oplus F \, , \qquad F\sim \mathrm{Bern}(q) \, .
\end{align}
Equivalently, trusted randomization can be absorbed into the very definition of Bob's measurement. 
This measurement is then characterized by the POVM elements
\begin{align}
\Lambda_0 & = (1-q) Z_0 + q Z_1 \, , \\
\Lambda_1 & = (1-q) Z_1 + q Z_0 \, ,
\end{align}
where $Z_j$ ($j=0,1$) are the $\mathbb{Z}$ eigenprojectors.

Following Ref.~\cite{winick2018reliable}, we introduce the \textit{key map} $\mathcal{G}$ associated to such a measurement.
{It is convenient to define first the coherent key-writing isometry}
\begin{align}
{K = \sum_{j=0,1} \mathbb{I}_A\otimes \sqrt{\Lambda_j}\otimes |j\rangle_{Y_1},}
\label{eq:key_isometry}
\end{align}
{which satisfies $K^\dagger K=\mathbb I_{AB}$ because $\Lambda_0+\Lambda_1=\mathbb I_B$. The key map is therefore}
\begin{align}
{\mathcal G(\rho_{AB})=K\rho_{AB}K^\dagger .}
\label{eq:key_map_coherent}
\end{align}
The dependence of $K$ and $\mathcal G$ on the trusted flipping probability $q$ is understood through the POVM elements $\Lambda_j$ and will not be displayed explicitly in the notation.

The analogous construction can be introduced for Bob's measurement in the $\mathbb{X}$ basis. The error-event projectors are
\begin{align}
\Pi_X^{\text{err}}&=\ket{\texttt{+-}}_{AB}\bra{\texttt{+-}}+
\ket{\texttt{-+}}_{AB}\bra{\texttt{-+}},\\
\Pi_Z^{\text{err}}&=\ket{01}_{AB}\bra{01}+\ket{10}_{AB}\bra{10}.
\end{align}
\begin{align}
Q_{\mathbb X}(\rho)&=\operatorname{tr}\!\left[\Pi_X^{\text{err}}\rho_{AB}\right],\qquad
Q_{\mathbb Z}(\rho)=\operatorname{tr}\!\left[\Pi_Z^{\text{err}}\rho_{AB}\right].
\label{subb1}
\end{align}
{For simplicity, in the main text we assume a symmetric channel, $Q_{\mathbb Z}(\rho)=Q_{\mathbb X}(\rho)=p_{\rm true}$; the value of $p_{\rm true}$ is inferred from the finite samples described next. See appendix \ref{app:asymmetric_qber} for the case $Q_{\mathbb Z}\neq Q_{\mathbb X}$.}

\subsection{{Parameter estimation}}
{Let $\widehat Q_{\mathbb{X}}$ and $\widehat Q_{\mathbb{Z}}$ denote the empirical error frequencies in the revealed test samples. 
For $b\in\{\mathbb X,\mathbb Z\}$, a two-sided Hoeffding bound gives}
\begin{align}
{\Pr\!\left[|\widehat Q_b-Q_b|\ge\delta_b\right]\le 2e^{-2k_b\delta_b^2},
\qquad k_b=k_\mathbb X\lor k_\mathbb Z.}
\end{align}
{We reveal a total fraction $f$ of the $N$ sifted rounds, split equally between the two bases, so that $k_b=fN/2$ and $m=(1-f)N$ rounds remain for key generation. We require $fN/2$ to be an integer. Splitting the parameter-estimation failure probability as $\epsilon_X=\epsilon_Z=\epsilon_{\rm PE}/2$ gives ($\delta=\delta_\mathbb X=\delta_\mathbb Z$)}
\begin{align}
{\delta=\sqrt{\frac{\ln(4/\epsilon_{\rm PE})}{fN}}\propto \frac{1}{\sqrt N}.}
\end{align}
{Under the symmetric-channel model, the confidence intervals are intersected to obtain}
{
\begin{align}
p_- &= \max\!\left\{0,\widehat Q_{\mathbb{X}}-\delta,\widehat Q_{\mathbb{Z}}-\delta\right\},\nonumber\\
p_+ &= \min\!\left\{\frac12,\widehat Q_{\mathbb{X}}+\delta,\widehat Q_{\mathbb{Z}}+\delta\right\}.
\label{eq:p_interval}
\end{align}
}
{The set of admissible bipartite states is thus defined as}
{
\begin{equation}
\begin{aligned}
\mathcal{S}_{\mathrm{PE}} := \Bigl\{\, &
\rho_{AB} \in \mathcal{D}(AB) \mid
\operatorname{tr}_B \rho_{AB}
= \frac{\mathbb{I}_A}{2},\\
& Q_{\mathbb{X}}(\rho_{AB})
= Q_{\mathbb{Z}}(\rho_{AB}),\\
& p_- \leq Q_{\mathbb{X}}(\rho_{AB}) \leq p_+ \,\Bigr\},
\end{aligned}
\label{eq:PE_constraints}
\end{equation}
where $\mathcal D(AB)$ indicated the set of quantum states on $AB$.
}
{If $p_->p_+$, the protocol aborts because the observed data are incompatible with the symmetric model at the chosen confidence level. In the numerical results we set $\widehat Q_{\mathbb{X}}=\widehat Q_{\mathbb{Z}}=p$.}

\subsection{Secret-key rate and R\'enyi-divergence formulation}

In order to assess the security of the protocol, it is convenient to introduce the isometry
\begin{align}
V_{Y_1\to YY_1}=\sum_{j=0,1}|j\rangle_Y\otimes|j\rangle_{Y_1}\langle j|_{Y_1} 
\end{align}
(which satisfies the identity $V^\dagger V=\mathbb{I}_{Y_1}$).
Formally, this creates a copy of $Y_1$ into an auxiliary register $Y$.
Including this latter isometry, we can write a five-partite pure state
\begin{align}
\rho_{ABYY_1E}=(V\otimes\mathbb I_E)\, \mathcal{G}(\rho_{ABE})\, (V^\dagger\otimes\mathbb I_E) \, .
\end{align}

We conclude this section noticing that the operation of partial trace, when applied on the auxiliary system $Y$, leaves the system in a state that is classical on system $Y_1$. We have
\begin{align}
\mathrm{tr}_{Y}( \rho_{ABYY_1E})
= \mathcal{Z} ( \mathcal{G} (\rho_{ABE}) ) \, ,   
\end{align}
where $\mathcal{Z}$ is the pinching map,
which fully dephases the register $Y_1$:
\begin{align}
\mathcal{Z}( \mathcal{G} (\rho_{ABE}) ) = \sum_{j=0,1} |j\rangle_{Y_1}\langle j| \,
\mathcal{G} (\rho_{ABE}) \, |j\rangle_{Y_1}\langle j| \, .
\end{align}

After sifting, {and after revealing the test samples,} Alice and Bob hold two highly correlated $m$-bit strings; we denote Bob's raw key as $Y^m$.
The leftover-hash lemma, based on R\'enyi entropy~\cite{dupuis2023privacy}, bounds the number $\ell$ of bits that can be extracted from $Y^m$ such that Eve has little or no information about them.
{Here $\lambda_{\rm EC}$ denotes the total public information disclosed for error correction and verification, whose two contributions are specified below. If}
\begin{align}
{
\ell \le \tilde H^\uparrow_\alpha(Y^m|E^m)-g_{\epsilon_{\rm PA}}(\alpha)-\lambda_{\rm EC},
}
\label{eq:keylength_new}
\end{align}
{
then one can extract a secret key with secrecy failure probability controlled by $\epsilon_{\rm PA}$, in addition to the parameter-estimation failure probability $\epsilon_{\rm PE}$.}
The above bound holds for any $\alpha\in(1,2]$, where
\begin{equation}
{g_{\epsilon_{\rm PA}}(\alpha)=\frac{\alpha}{\alpha-1}\log\!\left(\frac{1}{\epsilon_{\rm PA}}\right)-2}
\end{equation}
(we put $\log\equiv\log_2$), and $\tilde H^\uparrow_\alpha(A|B)$ denotes the optimized sandwiched R\'enyi conditional entropy,
\begin{align}
\tilde H^\uparrow_\alpha(A|B)_\rho
=\sup_{\sigma_B}\frac{1}{1-\alpha}\log
\mathrm{tr}\!\left[
\left(
\sigma_B^{\frac{1-\alpha}{2\alpha}}\,
\rho_{AB}\,
\sigma_B^{\frac{1-\alpha}{2\alpha}}
\right)^\alpha
\right] 
\end{align}
(we put $\sigma_B = \mathbb I_A\otimes \sigma_B$).
It is convenient to recall the definition of the
sandwiched R\'enyi divergence \cite{tomamichel2015quantum, tomamichel2014relating}
\begin{align}
\tilde D_\alpha(\rho\|\omega)
:=\frac{1}{\alpha-1}\log
\mathrm{tr}\!\left[
\left(
\omega^{\frac{1-\alpha}{2\alpha}}\;
\rho\;
\omega^{\frac{1-\alpha}{2\alpha}}
\right)^\alpha
\right] \, ,
\end{align}
and its relation to the conditional entropy,
\begin{align}
\tilde H^\uparrow_\alpha(A|B)_\rho
=-\inf_{\sigma_B}\tilde D_\alpha\,\big(\rho_{AB}\,\big\|\,\mathbb{I}_A\otimes\sigma_B\big) \, .
\end{align}
{The syndrome leakage is evaluated conservatively at the largest QBER compatible with the confidence region. Since $s(p,q)=q+(1-2q)p$ and $h_2(s)$ is nondecreasing in $p$ on the domain considered, we set}
{
\begin{align}
\ell_{\rm leak}=m\,h_2(s(q)), \quad
s(q)=q+(1-2q)p_+.
\label{eq:leak_q}
\end{align}
}
{Let $K_A$ and $K_B$ denote Alice's and Bob's reconciled keys before verification, and let $\Omega_{\mathrm{acc}}$ denote the acceptance event that their verification tags agree. The two-universal verification hash uses a tag of length}
\begin{align}
{t_{\rm EC}=\left\lceil\log_2\frac{1}{\epsilon_{\rm EC}}\right\rceil,}
\label{eq:tEC}
\end{align}
{which ensures $\Pr[K_A\ne K_B\wedge\Omega_{\mathrm{acc}}]\le\epsilon_{\rm EC}$. We assume ideal Slepian--Wolf error correction, corresponding to unit reconciliation inefficiency. The total information disclosed for reconciliation and verification is then}
\begin{align}
{\lambda_{\rm EC}:=\ell_{\rm leak}+t_{\rm EC}=m\,h_2(s(q))+t_{\rm EC}.}
\label{eq:lambdaEC}
\end{align}

Using the duality relation between sandwiched R\'enyi entropies~\cite{tomamichel2014relating}, and within the framework collective attacks, we obtain
\begin{align}
\tilde H^\uparrow_\alpha(Y^m|E^m)
=
-m\,\tilde H^\uparrow_{\beta}(Y|ABY_1) \, ,
\qquad \beta=\frac{\alpha}{2\alpha-1} \, ,
\end{align}
which holds true since $\rho^{\otimes m}_{ABYY_1E}$ is a pure state and is a tensor power.
Moreover, the conditional entropy on the right-hand side admits a divergence representation directly in terms of the key map and pinching map introduced in the previous section~\cite{staffieri2026finiteCV}:
\begin{align}
-\tilde H^\uparrow_{\beta}(Y|ABY_1)
=
\inf_{\sigma_{ABY_1}}
\tilde D_{\beta}\,\Big(\mathcal{G}(\rho_{AB})\ \Big\|\ \mathcal{Z}(\sigma_{ABY_1})\Big) \, .
\end{align}
{The privacy term is obtained by minimizing over all states compatible with the observed data. The bounds $p_-$ and $p_+$ are determined by the observed QBER $p$ and Eq.~\eqref{eq:p_interval}. We write its per-key-round value as}
{
\begin{align}
d_\beta(q;N,f,p):=
\min_{\substack{\rho_{AB},\sigma_{ABY_1}\\
\rho_{AB}\in \mathcal S_{\mathrm {PE}}}}
\tilde D_\beta\!\Big(\mathcal G(\rho_{AB})\,\Big\|\,\mathcal Z(\sigma_{ABY_1})\Big).
\label{eq:dbeta}
\end{align}
}
{The secret-key rate is normalized to the $N$ sifted rounds before test samples are revealed. Using $m=(1-f)N$ and the total leakage $\lambda_{\rm EC}$ gives}
{
\begin{align}
r(&N,\alpha,p,q;f)
=\frac{m\,d_\beta(q;N,f,p)-g_{\epsilon_{\rm PA}}(\alpha)-\lambda_{\rm EC}}{N}
\nonumber\\
&=(1-f)\!\left[d_\beta(q;N,f,p)-h_2(s(q))\right]
-\frac{g_{\epsilon_{\rm PA}}(\alpha)+t_{\rm EC}}{N}.
\label{eq:rate_div_form}
\end{align}
}
{The factors of $N$ in $m=(1-f)N$ and $\ell_{\rm leak}=m h_2(s)$ cancel against the rate denominator. The finite-key security bound is composable: parameter-estimation failure (with probability at most $\epsilon_{\rm PE}$) and privacy-amplification failure (at most $\epsilon_{\rm PA}$) jointly contribute at most $\epsilon_{\rm PE}+\epsilon_{\rm PA}$ to the secrecy error. Error verification separately ensures correctness except with probability at most $\epsilon_{\rm EC}$. The total security parameter is therefore bounded by $\epsilon_{\rm tot}\le \epsilon_{\rm PE}+\epsilon_{\rm PA}+\epsilon_{\rm EC}$.}

\subsection{{Conic formulation and numerical optimization}}
{For fixed $(N,f,q,\alpha)$, the inner minimization in Eq.~\eqref{eq:dbeta} is formulated using the R\'enyiQKD cone introduced in Ref.~\cite{navarro2025finite}. Let $h$ denote the scalar conic variable associated with the facial-reduced R\'enyi trace functional. In the regime $\beta\in[2/3,1)$ relevant here, the formulation of the numerical optimization follows}
{
\begin{align}
h^\star_\beta(q;N,f,p)
=&\min_{h,\rho,\sigma}\quad h
\nonumber\\[-1mm]
\text{s.t.}\quad
&\mathrm{tr}_B\rho=\frac{\mathbb I_A}{2},\qquad \mathrm{tr}\sigma=1,\nonumber\\
&Q_{\mathbb{X}}(\rho)=Q_{\mathbb{Z}}(\rho),
\,p_-\le Q_{\mathbb{X}}(\rho)\le p_+,\nonumber\\
&(h,\rho,\sigma)\in
\mathcal K_{\mathrm{R\acute enyiQKD}}^{\beta,\mathrm{id},\mathcal Z,K}.
\label{eq:conic_problem}
\end{align}
}
{We emphasize that the conic program in
Eq.~\eqref{eq:conic_problem} optimizes the R\'enyi trace functional
rather than its logarithm. This is sufficient because, for fixed
$(N,f,q,\alpha)$, the logarithm is strictly monotonic and
$1/(\beta-1)<0$. Hence minimizing the sandwiched R\'enyi divergence is
equivalent to maximizing the corresponding trace functional, or,
equivalently, to minimizing the conic scalar $h$. The logarithm is
therefore applied only after the conic optimum has been obtained, and no
additional logarithm cone is required for the inner optimization
considered here.}
{The privacy term is recovered from the optimal conic scalar as}
\begin{align}
{d_\beta(q;N,f,p)
=\frac{\log_2[-h^\star_\beta(q;N,f,p)]}{\beta-1}.}
\label{eq:conic_to_entropy}
\end{align}
{The facial reduction entering Eq.~\eqref{eq:conic_problem} is particularly simple because the key map is generated by the isometry $K$: $\widehat{\mathcal G}=\mathrm{id}$, $\widehat{\mathcal Z}=\mathcal Z$, and $S=K$. The derivation and the definition of the cone are given in Appendix~\ref{app:conic}.}

{The final rate is obtained through the outer optimization}
{
\begin{align}
r^\star(N,p)=
\max_{\substack{0<f<1\\0\le q\le1/2\\1<\alpha\le2}}
r(N,\alpha,p,q;f).
\label{eq:global_opt}
\end{align}
}
{The bounds $p_\pm$ depend on both $N$ and $f$, so the feasible set and the optimal $(f,q,\alpha)$ generally change with the block size; the outer optimization is therefore repeated for every $N$. Numerically we use ConicQKD.jl~\cite{ConicQKDJulia} with Hypatia. Close to the von Neumann limit, where the transformation in Eq.~\eqref{eq:conic_to_entropy} amplifies finite-precision errors, the relevant points are recomputed in extended precision. Numerical accuracy is assessed through the primal--dual conic gap and feasibility residuals; these diagnostics provide a numerical consistency check at solver precision rather than an exact-arithmetic rigorous certificate.}

\section{Results}
\label{sec:results}

{In this section we present our results assessing the effect of preprocessing
noise on composable security in the finite-size regime. Throughout all numerical analyses, we adopt a total security budget $\epsilon_{\rm tot}=10^{-10}$, equally allocated as $\epsilon_{\rm PE}=\epsilon_{\rm PA}=\epsilon_{\rm EC}=\epsilon_{\rm tot}/3$.
First, we determine in which interval of values of $\alpha$ adding trusted
noise improves the security of the protocol.
For this analysis we assume that the QBER is known exactly, thus neglecting
the statistical penalty associated with parameter estimation. This allows us
to isolate the advantage provided by trusted preprocessing. In particular we consider}
{
\begin{align}
    \Delta r(\alpha)
    =
    \max_p
    \left[
    r(\alpha,p,q^*)-r(\alpha,p,q=0)
    \right].
    \label{Delta_r}
\end{align}
}
{Under the exact-QBER assumption, all additive finite-size corrections
which are independent of $q$ cancel in taking the difference between the two
rates. Therefore, apart from the statistical penalty associated with
parameter estimation, this quantity is independent of the block size and
shows directly that the advantage due to trusted noise persists in the
finite-size regime.
In taking the maximum with respect to the disturbance parameter $p$, we identify
the QBER values for which trusted bit flipping is most beneficial.}

\begin{figure}[t]
    \centering
    \includegraphics[width=\linewidth]{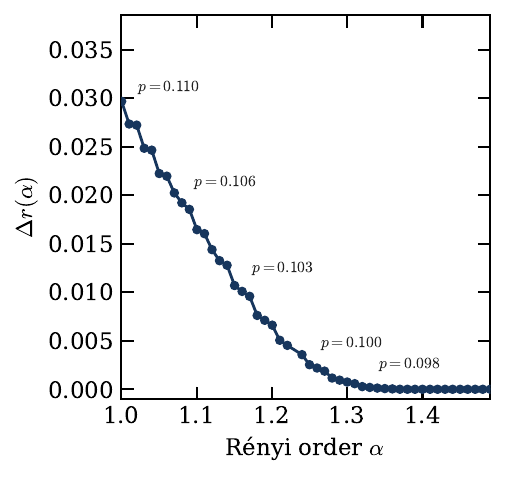}
    \caption{\justifying
    {Largest difference between the secret-key rate obtained with the
    optimal trusted-noise level $r(\alpha,p,q^*)$ and that obtained without
    trusted randomization, $r(\alpha,p,q=0)$, as in
    Eq.~\eqref{Delta_r}, plotted versus the R\'enyi order $\alpha$.
    For this figure the QBER is assumed to be known exactly and the
    parameter-estimation penalty is therefore omitted. The value of $p$ at
    which $\Delta r$ is maximized is displayed above the corresponding data
    points.}}
    \label{fig:DeltaR}
\end{figure}

{In Fig.~\ref{fig:DeltaR}, the above quantifier is plotted against the
R\'enyi order $\alpha$.
We observe that when $\alpha\to1^+$ one obtains the highest gain from local
randomization, in particular, the largest enhancement is found close to $p=0.11$, which is
the well-known threshold value for BB84 in the absence of preprocessing in the asymptotic limit $N\to \infty$, 
noise~\cite{shor2000simple}.
At the smallest R\'enyi order considered, $\alpha=1.0005$, the conic
optimization gives $\Delta r\simeq 0.0297$, with the maximum
attained at $p=0.110$.
As $\alpha$ increases, the gain decreases, while the QBER value for which
trusted preprocessing is most advantageous remains in the high-noise region.
This expresses one key feature already observed in the original analysis:
when trusted noise is beneficial, the most significant improvement of the
key rate occurs close to the maximum QBER tolerated without preprocessing.
Finally, around $\alpha\simeq1.4$, the addition of trusted bit flipping
ceases to improve the key rate, and the optimal solution becomes $q^*=0$.
As shown in Appendix~\ref{trust_min_ent}, in the limit
$\alpha\to\infty$ trusted preprocessing is never beneficial.}

\begin{figure}[t]
    \centering
    \includegraphics[width=\linewidth]{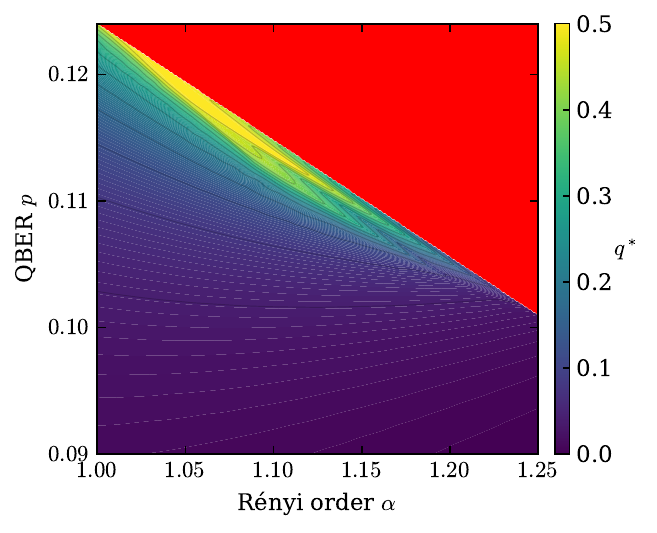}
    \caption{\justifying
    {Optimal trusted-noise level $q^*(\alpha,p)$ as a function of the
    QBER $p$ and the R\'enyi order $\alpha$. The $(\alpha,p)$ domain is
    restricted to those points for which a positive secret-key rate can be
    obtained; the forbidden domain is shown in red. The optimal trusted
    flipping probability increases as the disturbance approaches the boundary
    of the positive-key-rate region.}}
    \label{3D}
\end{figure}

{The heat map in Fig.~\ref{3D} shows the trend of the optimal value of
the trusted noise $q^*(\alpha,p)$ as a function of $\alpha$ and $p$.
Here we restrict the domain of $(\alpha,p)$ in order to consider only
non-vanishing key-rate values.
As expected from the analysis of $\Delta r(\alpha)$, for small values of the
disturbance the optimal level of trusted randomization is nearly zero.
The trend changes when $p$ approaches the region $p\simeq0.10$--$0.11$:
trusted preprocessing becomes increasingly relevant and the optimal flipping
probability grows as one approaches the boundary of the positive key-rate
region.
Clearly, $q^*$ also increases with the R\'enyi order close to this boundary,
until the secret-key rate eventually vanishes.}

\begin{figure}[t]
    \centering
    \includegraphics[width=\linewidth]{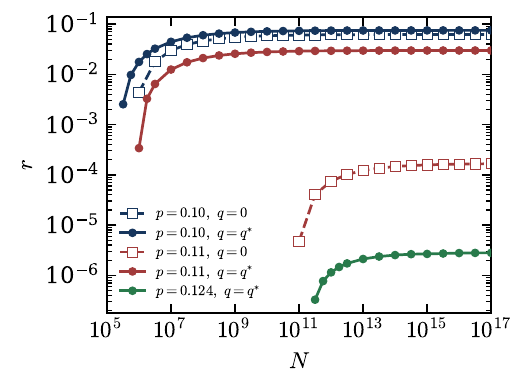}
    \caption{\justifying
    {Key rates at finite block size including finite-sample parameter
    estimation. We compare the rate obtained without trusted noise with that
    achieved using the optimal trusted-noise probability $q^*$. For every
    block size the parameter-estimation fraction $f$ and the R\'enyi order
    $\alpha$ are optimized, together with $q$ for the trusted-noise branch.
    Different curves refer to $p=0.10$, $0.11$, and $0.124$.
    The total security parameter is $\epsilon_{\rm tot}=10^{-10}$, with $\epsilon_{\rm PE}=\epsilon_{\rm PA}=\epsilon_{\rm EC}=\epsilon_{\rm tot}/3$. As the QBER increases, trusted noise becomes
    increasingly beneficial both in terms of the secret-key rate and of the
    minimum block length required for non-zero key generation.}}
    \label{fig:confronto_rate}
\end{figure}

{In Fig.~\ref{fig:confronto_rate} we show the best finite-size secret-key
rate estimates versus the sifted block length $N$ for fixed values of the
observed QBER $p$.
Unlike Figs.~\ref{fig:DeltaR} and~\ref{3D}, parameter estimation is now
explicitly included. For each $N$ and $p$, the rate for the
trusted-noise branch is optimized over the
test fraction $f$, the R\'enyi order $\alpha$, and $q$.
For $p=0.10$, Fig.~\ref{fig:confronto_rate} indicates that preprocessing is
required to obtain a non-zero key rate at the smallest block sizes shown.
The optimized trusted-noise branch becomes positive around
$N\simeq3.2\times10^5$, whereas the $q=0$ branch becomes positive on the
sampled grid at $N=10^6$.
At $N=10^6$, the optimized trusted-noise rate is
$r\simeq1.76\times10^{-2}$, obtained for
$f\simeq0.348$, $q^*\simeq0.072$, and $\alpha^*\simeq1.013$.
For larger block sizes the enhancement approaches a nearly
block-size-independent value, as finite-size and parameter-estimation
penalties become progressively less important.}

{For $p=0.11$, trusted local randomization yields a much more substantial
improvement.
Most notably, trusted noise reduces by several orders of magnitude the
minimum block length required to extract a secret key.
The optimized branch already gives a positive rate at $N=10^6$, where
$r\simeq3.4\times10^{-4}$ for
$f\simeq0.565$, $q^*\simeq0.183$, and $\alpha^*\simeq1.020$.
By contrast, without trusted noise no positive rate is obtained on the
sampled grid before $N\simeq10^{11}$.
As $N$ increases, the optimal R\'enyi order approaches $\alpha=1$, as
expected, and the two branches progressively approach their asymptotic
behavior.}

{Finally, for QBER values above the standard BB84 threshold, secrecy can
be attained only with the assistance of trusted noise.
The curve at $p=0.124$ illustrates this regime.
A positive key rate appears only for much larger block sizes, with the
transition occurring between approximately $1.8\times10^{11}$ and
$3.2\times10^{11}$ in the sampled data, whereas the $q=0$ branch cannot
generate a secret key at this QBER.
The achievable rate remains small even at very large $N$, consistently with
the fact that $p=0.124$ lies very close to the asymptotic trusted-noise
threshold.}

\begin{figure}[t]
    \centering
    \includegraphics[width=\linewidth]{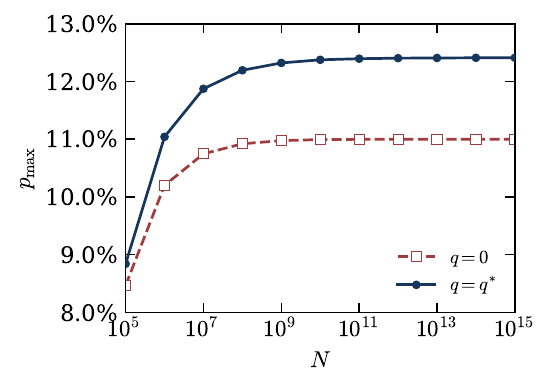}
    \caption{\justifying
    {Maximum tolerable QBER $p_{\rm max}$ plotted versus the sifted block
    length $N$. The two curves compare the threshold obtained with optimal
    trusted noise and without trusted noise, including finite-sample parameter
    estimation. The test fraction $f$ is optimized independently for the two
    branches and the total security parameter is fixed at
    $\epsilon_{\rm tot}=10^{-10}$, with $\epsilon_{\rm PE}=\epsilon_{\rm PA}=\epsilon_{\rm EC}=\epsilon_{\rm tot}/3$. Trusted preprocessing consistently increases
    the tolerable QBER, while for large block sizes the two curves approach
    the asymptotic values $p_{\rm max}\simeq0.124$ and
    $p_{\rm max}\simeq0.11$.}}
    \label{maxqber}
\end{figure}

{Finally, we determine the maximum tolerable QBER level, denoted
$p_{\rm max}$, as a function of the finite block size, both with and without
trusted noise.
Figure~\ref{maxqber} shows that preprocessing becomes increasingly beneficial
as the block size grows.
At $N=10^5$, the largest tolerable QBER is approximately $0.0884$ with
trusted preprocessing and $0.0847$ without it.
Already at $N=10^6$, these values increase to approximately $0.1104$ and
$0.1020$, respectively, while at $N=10^7$ they reach approximately $0.1187$
and $0.1075$.
For larger block lengths, trusted local randomization always guarantees a
higher QBER tolerance.
The two thresholds progressively saturate at
$p_{\rm max}\simeq0.124$ and $p_{\rm max}\simeq0.11$, respectively with
and without trusted preprocessing.}

\paragraph*{\normalfont \textbf{Numerical benchmark.}}

{As a numerical benchmark, we compare the ConicQKD.jl implementation~\cite{ConicQKDJulia} with the
independent two-variable Frank--Wolfe implementation described in
Appendix~\ref{gradient}. Both methods solve the same fixed-QBER optimization
problem at identical $(p,\alpha,q)$ values. The matched instances are chosen
to probe the numerical regimes most relevant to the analysis: points on the
optimized-noise branch, points on the $q=0$ branch at different R\'enyi orders,
and points close to the secrecy threshold. The latter also include cases close
to the von Neumann limit, where the key rate is small and numerical accuracy
is particularly important. The converged Double64 conic result is used as the
numerical reference in Table~\ref{tab:fw_benchmark}.}

\begin{table}[t]
\centering
\begingroup

\caption{\justifying
Comparison between the conic and Frank--Wolfe implementations on the same
fixed-$(p,\alpha,q)$ BB84 optimization problem. The reported ranges refer to
the matched benchmark points.}
\label{tab:fw_benchmark}
\scriptsize
\setlength{\tabcolsep}{2.5pt}
\resizebox{\columnwidth}{!}{
\begin{tabular}{lccc}
\hline
Method & Iterations & Runtime (s) & $|r-r_{\rm ref}|$ \\
\hline
Conic, Float64
& $17$--$61$
& $0.014$--$0.064$
& $\le 1.9\times10^{-9}$ \\

Conic, Double64
& $20$--$31$
& $0.150$--$0.536$
& reference \\

FW, 500 it.
& $500$
& $13.6$--$19.6$
& $1.7$--$5.3\times10^{-3}$ \\

FW, 5000 it.
& $5000$
& $139$--$196$
& $5.1\times10^{-6}$--$4.5\times10^{-4}$ \\

FW, up to 20000 it.
& $9813$--$20000$
& $375$--$774$
& $2.1\times10^{-6}$--$1.4\times10^{-4}$ \\
\hline
\end{tabular}}
\endgroup
\end{table}

{The benchmark shows that the conic formulation reaches the converged
Double64 reference within a few tens of interior-point iterations, whereas
the Frank--Wolfe convergence is substantially slower. In Float64, the conic
results agree with the Double64 reference to within $1.9\times10^{-9}$ over
the matched benchmark points, with runtimes of order $10^{-2}$~s. By
comparison, the 500-iteration Frank--Wolfe runs require 13.6--19.6~s while
the remaining deviation from the reference is still of order $10^{-3}$.
Increasing the iteration budget to 5000 reduces this deviation to the
$10^{-6}$--$10^{-4}$ range, but requires 139--196~s, while the longest runs
require several minutes per fixed $(p,\alpha,q)$ value. The direct comparison
also includes positive-rate near-threshold instances at
$\alpha-1=1.5\times10^{-4}$ and $\alpha-1=5.6\times10^{-4}$, confirming that
the slower Frank--Wolfe convergence persists in the regime close to the von
Neumann limit that is most relevant for threshold calculations. These results
motivate the use of the conic formulation for the production results, while
the two-variable Frank--Wolfe implementation is retained as an independent
numerical cross-check.}

\section{Conclusions}

Preprocessing such as the injection of trusted noise is known to enhance the performance of QKD protocols, especially in terms tolerated QBER.
This phenomenon has been extensively studied in the asymptotic limit in the number of channel uses~\cite{renner2005information,kraus2005lower,renes2007noisy,mertz2013quantum}.
{Here we have studied trusted local randomization at finite block size under collective attacks using optimized sandwiched R\'enyi entropies. The security bound retains the full variational optimization over both the bipartite state and the auxiliary state entering the optimized conditional entropy.}
{The main numerical tool is the nonsymmetric R\'enyiQKD conic formulation of Ref.~\cite{navarro2025finite}. For the trusted-noise key map the facial reduction is analytic, which allows the variational problem to be solved directly while retaining the physical and statistical constraints on the underlying state. The trusted flipping probability and the R\'enyi order are then optimized externally.}
{An independent two-variable Frank--Wolfe formulation is available as a cross-check of the same variational quantity.}
{The resulting finite-size analysis quantifies the regime in which trusted preprocessing improves the secret-key rate and the corresponding increase in the maximum tolerable QBER.}

In terms of applications, our analysis may be extended to the framework of both \emph{Device-Independent} QSS \cite{zhang2024device}  and \emph{Device-Independent} QKD~\cite{Ghoreishi2025a,tan2022improved,ulu2025device}.
Furthermore, tighter estimates on the key rates may be obtained using the latest leftover-hash lemma \cite{regula2026rethinking}, which introduces measured smooth entropies and a novel Hermitian smoothing technique. This approach allows for smoothing over non-positive operators, providing a more precise one-shot characterization of privacy amplification and recovering the sharpest achievability results currently available. 
{A natural direction for future work is to extend our security
analysis from collective to general coherent attacks using the
marginal-constrained entropy accumulation theorem (MEAT),
leveraging the R\'enyi-entropy conic optimization framework
developed in Refs.
\cite{navarro2025finite,Navarro2026DMQKD}.}
{Another interesting direction is the extension of trusted noisy
preprocessing to blocks of raw-key bits, allowing for correlated
noise distributions $P(F_1,F_2)$ and possibly combining them
with structured blockwise processing, as motivated by previous
studies of noisy preprocessing and advantage distillation
\cite{SmithRenesSmolin2008,KernRenes2008,GottesmanLo2003},
to investigate their potential benefits within our finite-size
R\'enyi framework.}

The Frank--Wolfe implementation used in this work is available in Ref.~\cite{GG84}, while the code for the conic computations is provided in Ref.~\cite{TrustedNoiseQKD2026} and builds on
ConicQKD.jl~\cite{ConicQKDJulia}.

\begin{acknowledgments}
This work has received support by the European Union's Horizon Europe research and innovation programme under the Project ``Quantum Secure Networks Partnership'' (QSNP, Grant Agreement No.~101114043),
and by INFN through the project ``QUANTUM''.
\end{acknowledgments}

\appendix

\section{{Conic representation and facial reduction}}
\label{app:conic}

{We provide here the details underlying Eq.~\eqref{eq:conic_problem}. Define the facial-reduced R\'enyi trace functional}
\begin{align}
{\widehat\Psi_\beta(\rho,\sigma)
=
\left\|
\widehat{\mathcal Z}(\sigma)^{\frac{1-\beta}{2\beta}}
S\,\widehat{\mathcal G}(\rho)^{1/2}
\right\|_{2\beta}^{2\beta}.}
\label{eq:psi_hat_app}
\end{align}
{For $\beta<1$, $\widehat\Psi_\beta$ is jointly concave in the relevant domain. Since $s_\beta=\mathrm{sign}(\beta-1)=-1$, the function $s_\beta\widehat\Psi_\beta=-\widehat\Psi_\beta$ is convex and positively homogeneous of degree one. Its epigraph therefore defines the R\'enyiQKD cone}
\begin{align}
{\mathcal K_{\mathrm{R\acute enyiQKD}}^{\beta,\widehat{\mathcal G},\widehat{\mathcal Z},S}
=
\mathrm{cl}\left\{(h,\rho,\sigma):
 h\ge -\widehat\Psi_\beta(\rho,\sigma)\right\}.}
\label{eq:renyi_cone_app}
\end{align}
{Minimizing $h$ over this cone is equivalent to maximizing $\widehat\Psi_\beta$. Because $1/(\beta-1)<0$, this is in turn equivalent to minimizing the sandwiched R\'enyi divergence; the latter is recovered as $\log_2(-h^\star)/(\beta-1)$, as used in Eq.~\eqref{eq:conic_to_entropy}.}

{For the trusted-noise key map, the facial reduction is analytic. From Eq.~\eqref{eq:key_isometry}, $K^\dagger K=\mathbb I_4$ and}
\begin{align}
{\mathcal G(\rho)=K\rho K^\dagger.}
\end{align}
{Although $\mathcal G(\rho)$ is represented on the eight-dimensional space $ABY_1$, its support is contained in the four-dimensional image of $K$. Hence we may choose}
\begin{align}
{W_G=K,\qquad \widehat{\mathcal G}=\mathrm{id}.}
\end{align}
{For the pinching map no analogous support reduction is required in the present formulation, so that}
\begin{align}
{W_Z=\mathbb I_8,\qquad \widehat{\mathcal Z}=\mathcal Z.}
\end{align}
{The isometry entering the facial-reduced R\'enyi functional is therefore}
\begin{align}
{S=W_Z^\dagger W_G=K.}
\end{align}
{This yields exactly the cone appearing in Eq.~\eqref{eq:conic_problem}. The additional affine constraints only restrict the admissible $\rho_{AB}$ and do not modify the R\'enyiQKD cone itself.}

\section{{Two-variable Frank--Wolfe formulation and gradient of the sandwiched R\'enyi divergence}}

\label{gradient}

{This appendix describes the independent FW implementation used for the fixed-QBER benchmark of Sec.~\ref{sec:results}, where the constraints are $Q_{\mathbb{X}}(\rho)=Q_{\mathbb{Z}}(\rho)=p$.}

Here we derive the relation for the gradients $\nabla_\rho$ and $\nabla_\sigma$ 
{used in the independent FW benchmark}. The derivation follows a similar strategy to that of Ref.~\cite{chung2025generalized}, but extends it to the present setting where the objective functional depends on two distinct variables, $\rho$ and $\sigma$, rather than on a single optimization variable. We consider the objective 
\begin{align}
    D_\beta(X||Y) = \frac{1}{\beta-1}\log(Q), \qquad Q\equiv Q_\beta(X||Y)=\mathrm{tr}(\Xi^\beta);
\end{align}
where we recall: $X=\mathcal G(\rho_{AB})$, $Y=\mathcal Z(\sigma_{ABY_1})$, $\Xi = Y^\mu XY^\mu$, $\mu=(1-\beta)/2\beta$. Throughout this section we will denote with $\partial_{\Delta X}$ the directional derivative in such a way that $\forall \Delta X=\Delta X^\dagger$, \cite{bhatia2013matrix}
\begin{align}\label{gradientdef}
    \partial_{\Delta X}F(X) = \frac{d}{dt}F(X+t\Delta X)\big|_{t=0}=\mathrm{tr}\left((\nabla_XF)\Delta X\right).
\end{align}
We aim to compute the gradient with respect to $\rho_{AB}$. To do that, let us start form what follows. We need to fix $Y$ and choose a \emph{direction} $\Delta X$, then we have 
\begin{align}
    \partial_{\Delta X} \Xi = Y^\mu \Delta X Y^\mu.
\end{align}
Thus
\begin{align}
    \partial _{\Delta X} Q &= \beta\mathrm{tr}(\Xi ^{\beta-1} \partial_{\Delta X} \Xi) = \beta \mathrm{tr} (\Xi ^{\beta-1}Y^\mu \Delta X Y^\mu) \nonumber\\
    &= \beta\mathrm{tr}(Y^\mu\Xi ^{\beta-1}Y^\mu \Delta X )=\mathrm{tr}(\chi_2 \Delta X),
\end{align}
where we used cyclicity of the trace and $\chi_2=\beta\,Y^\mu\Xi ^{\beta-1}Y^\mu $. Hence the directional derivative of the Rényi divergence along $\Delta X$ is
\begin{align}
      \partial _{\Delta X} D_\beta(X\|Y)&
=\frac{1}{(\beta-1)\ln 2}\frac{1}{Q}\,
\partial_{\Delta X}Q \nonumber \\
&=\frac{1}{(\beta-1)\ln 2}
\mathrm{tr}
\left(
\frac{\chi_2}{Q}\,
\Delta X 
\right).
\end{align}
From Riesz lemma \eqref{gradientdef}, we can identify
\begin{align}
    \nabla_X D_\beta (X||Y) =\frac{1}{(\beta-1)\ln2}\frac{\chi_2}{Q}
\end{align}
Finally we observe that $X=\mathcal G(\rho_{AB}) \Longrightarrow \Delta X= \mathcal G(\Delta\rho)$ in such a way that
\begin{align}
\partial_{\Delta\rho}D_\beta\bigl(\mathcal G(\rho)\|Y\bigr)
&=\mathrm{tr}\bigl[
(\nabla_X D_\beta)\,\mathcal G(\Delta\rho)\bigr]\nonumber\\
&=\mathrm{tr}\bigl[
\mathcal G^\dagger(\nabla_X D_\beta)\,\Delta\rho
\bigr],
\end{align}
hence
\begin{align}
\nabla_{\rho}\,D_\beta\bigl(
\mathcal G(\rho_{AB})\|&\mathcal Z(\sigma_{ABY_1})
\bigr)
=\mathcal G^\dagger\!\bigl(\nabla_X D_\beta(X\|Y)\bigr)\nonumber\\
&=\frac{1}{(\beta-1)\ln2}\mathcal{G}^\dagger
\left(
\frac{\chi_2}{Q_\beta(X||Y)}
\right).
\label{eq:grad_rho_app}
\end{align}

Now we compute the gradient with respect to $\sigma_{ABY_1}$.
We fix $X$ and consider a direction $\Delta Y$, and consider the Fréchet derivative of $Y^\mu$ in the integral representation \cite{rubboli2022new}
\begin{equation}
\partial_{\Delta Y}Y^\mu
=L(\mu)\int_0^\infty (Y+s)^{-1}\Delta Y (Y+s)^{-1}s^\mu\,ds,
\label{eq:lemma_power_app}
\end{equation}
with $L(\mu)=\frac{\sin(\pi\mu)}{\pi}$. Define the linear map
\begin{equation}
\mathcal T_\mu^Y(\Delta Y):=\partial_{\Delta Y}Y^\mu.
\label{eq:T_def_app}
\end{equation}
Then
\begin{equation}
\partial_{\Delta Y}\Xi
=\mathcal T_\mu^Y(\Delta Y)\,X\,Y^\mu
+Y^\mu X\,\mathcal T_\mu^Y(\Delta Y).
\label{eq:partial_Xi_Y_app}
\end{equation}
in such a way that
\begin{align}
&\partial_{\Delta Y}Q
=\beta\,\mathrm{tr}\bigl(\Xi^{\beta-1}\,\partial_{\Delta Y}\Xi\bigr)\nonumber\\
&=\beta\,\mathrm{tr}\bigl(\Xi^{\beta-1}\mathcal T_\mu^Y(\Delta Y)\,X\,Y^\mu\bigr)
+\beta\,\mathrm{tr}\bigl(\Xi^{\beta-1}Y^\mu X\,\mathcal T_\mu^Y(\Delta Y)\bigr)\nonumber\\
&=\beta\,\mathrm{tr}\bigl(XY^\mu\Xi^{\beta-1}\,\mathcal T_\mu^Y(\Delta Y)\bigr)
+\beta\,\mathrm{tr}\bigl(\Xi^{\beta-1}Y^\mu X\,\mathcal T_\mu^Y(\Delta Y)\bigr).
\label{eq:partial_Q_Y_app1}
\end{align}
We next use the self-adjointness induced by \eqref{eq:lemma_power_app}:
for any $A,B$,
\begin{equation}
\mathrm{tr}\bigl(A\,\mathcal T_\mu^Y(B)\bigr)=\mathrm{tr}\bigl(\mathcal T_\mu^Y(A)\,B\bigr),
\label{eq:T_selfadj_app}
\end{equation}
which follows by inserting \eqref{eq:lemma_power_app} and using cyclicity of the trace. Setting
\begin{equation}
A_1=X\,Y^\mu\,\Xi^{\beta-1},
\qquad
A_3=\Xi^{\beta-1}Y^\mu X,
\label{eq:B13_defs_app}
\end{equation}
we can rewrite \eqref{eq:partial_Q_Y_app1} as
\begin{align}
\partial_{\Delta Y}Q
&=\beta\,\mathrm{tr}\bigl(A_1\,\mathcal T_\mu^Y(\Delta Y)\bigr)
+\beta\,\mathrm{tr}\bigl(A_3\,\mathcal T_\mu^Y(\Delta Y)\bigr)\nonumber\\
&=\mathrm{tr}\Bigl(\beta\bigl(\mathcal T_\mu^Y(A_1)+\mathcal T_\mu^Y(A_3)\bigr)\,\Delta Y\Bigr),
\label{eq:partial_Q_Y_app2}
\end{align}
where \eqref{eq:T_selfadj_app} is applied term--wise. Therefore,
\begin{align}
&\partial_{\Delta Y}D_\beta(X\|Y)
=\frac{1}{(\beta-1)\ln 2}\frac{1}{Q}\,\partial_{\Delta Y}Q \nonumber\\
&=\frac{1}{(\beta-1)\ln 2}
\mathrm{tr}
\left(
\frac{\beta\bigl(\mathcal T_\mu^Y(A_1)+\mathcal T_\mu^Y(A_3)\bigr)}{Q}\,
\Delta Y 
\right)\nonumber\\
&=\frac{1}{(\beta-1)\ln 2}
\left(
\frac{\chi_1+\chi_3}{Q}\,
\Delta Y 
\right)
\label{eq:partial_D_Y_app}
\end{align}
having set $\chi_1=\beta T_\mu^Y(A_1)$ and $\chi_3=\beta T_\mu^Y(A_3)$.
Comparing \eqref{eq:partial_D_Y_app} with \eqref{gradientdef}, we identify
\begin{equation}
\nabla_Y D_\beta(X\|Y)
=\frac{1}{(\beta-1)\ln 2}\,
\frac{\chi_1+\chi_3}{Q} .
\label{eq:grad_Y_app}
\end{equation}
Finally, $Y=\mathcal Z(\sigma_{ABY_1})$ implies $\Delta Y=\mathcal Z(\Delta\sigma)$, so
\begin{align}
\partial_{\Delta\sigma}D_\beta\!\bigl(X\|\mathcal Z(\sigma)\bigr)
&=\mathrm{tr}\bigl((\nabla_Y D_\beta)\,\mathcal Z(\Delta\sigma)\bigr)\nonumber\\
&=\mathrm{tr}\bigl(\mathcal Z^\dagger(\nabla_Y D_\beta)\,\Delta\sigma\bigr).
\end{align}
Hence
\begin{align}
\nabla_{\sigma}\,D_\beta\bigl(\mathcal G(\rho_{AB})\|&\mathcal Z(\sigma_{ABY_1})\bigr)
=\mathcal Z^\dagger\bigl(\nabla_Y D_\beta(X\|Y)\bigr)\nonumber\\
&=
\frac{1}{(\beta-1)\ln2}\mathcal{Z}
\left(
\frac{\chi_1+\chi_3}{Q_\beta(X||Y)}
\right)
\label{eq:grad_sigma_app}
\end{align}
which holds since for a pinching map $\mathcal Z^\dagger=\mathcal Z$.

{For a fixed-QBER benchmark, the two density operators $\rho$ and $\sigma$ are updated jointly through the block gradient $\nabla=\nabla_\rho\oplus\nabla_\sigma$. Each Frank--Wolfe step solves the linearized problem over the fixed-QBER feasible set.}

\section{{Extension to the six-state protocol}}
\label{app:sixstate}

{To assess whether the advantage provided by trusted local
randomization is specific to BB84, we extend the exact-QBER analysis of
Figs.~\ref{fig:DeltaR} and~\ref{3D} to the six-state protocol. In addition
to the $\mathbb Z$ and $\mathbb X$ bases, Alice and Bob now also use the
$\mathbb Y$ basis, while the raw key is still generated from the
$\mathbb Z$-basis outcomes and Bob applies the same trusted bit flip
introduced in Sec.~\ref{sec:method}, with probability $q$.}

{We denote the eigenstates of the Pauli-$Y$ operator by}
{
\begin{align}
    \ket{\texttt{+i}}
    &=
    \frac{\ket{0}+i\ket{1}}{\sqrt{2}},
    &
    \ket{\texttt{-i}}
    &=
    \frac{\ket{0}-i\ket{1}}{\sqrt{2}}.
\end{align}
}
{For the maximally entangled state $\ket{\Phi^+}$ used in the
entanglement-based representation, the outcomes in the $\mathbb Y$ basis
are ideally anticorrelated. Equivalently, after the conventional relabeling
of Bob's $\mathbb Y$-basis outcome, the error event is represented by the
projector}
{
\begin{align}
    \Pi_Y^{\mathrm{err}}
    =
    \ket{\texttt{+i},\texttt{+i}}
    \bra{\texttt{+i},\texttt{+i}}
    +
    \ket{\texttt{-i},\texttt{-i}}
    \bra{\texttt{-i},\texttt{-i}},
    \label{eq:Y_error_projector}
\end{align}
}
{and the corresponding QBER is}
{
\begin{align}
    Q_{\mathbb{Y}}(\rho)
    =
    \operatorname{tr}
    \left[
        \Pi_Y^{\mathrm{err}}\rho_{AB}
    \right].
    \label{eq:QY_sixstate}
\end{align}
}

{As in the BB84 analysis of Figs.~\ref{fig:DeltaR} and~\ref{3D},
we assume here that the disturbance is known exactly. The symmetric
six-state model is therefore specified by}
{
\begin{align}
    Q_{\mathbb{X}}(\rho)=Q_{\mathbb{Y}}(\rho)=Q_{\mathbb{Z}}(\rho)=p.
    \label{eq:sixstate_constraints}
\end{align}
}
{The trusted-noise key map and the R\'enyiQKD cone are unchanged.
The only modification of the inner optimization with respect to BB84 is
the additional affine constraint associated with $Q_{\mathbb{Y}}$. More explicitly,
the six-state conic problem reads}
{
\begin{align}
    h^{\star}_{\beta,6}(q,p)
    =&\min_{h,\rho,\sigma}\quad h
    \nonumber\\[-1mm]
    \text{s.t.}\quad
    &\operatorname{tr}_B\rho
    =\frac{\mathbb I_A}{2},
    \qquad
    \operatorname{tr}\sigma=1,
    \nonumber\\
    &Q_{\mathbb{X}}(\rho)=Q_{\mathbb{Y}}(\rho)=Q_{\mathbb{Z}}(\rho)=p,
    \nonumber\\
    &(h,\rho,\sigma)
    \in
    \mathcal K_{\mathrm{R\acute enyiQKD}}
    ^{\beta,\mathrm{id},\mathcal Z,K}.
    \label{eq:conic_problem_sixstate}
\end{align}
}
{Accordingly, the six-state privacy term is obtained from
$h^{\star}_{\beta,6}$ through the same relation as in
Eq.~\eqref{eq:conic_to_entropy}. Since the QBER is fixed exactly in this
analysis, no parameter-estimation confidence interval is introduced.}

{To quantify the advantage due specifically to trusted preprocessing,
we consider the six-state counterpart of Eq.~\eqref{Delta_r},}
{
\begin{align}
    \Delta r_{6}(\alpha)
    =
    \max_p
    \left[
        r_{6}(\alpha,p,q^*)
        -
        r_{6}(\alpha,p,q=0)
    \right],
    \label{eq:delta_sixstate}
\end{align}
}
{where $q^*$ is optimized independently at each $(\alpha,p)$. As in
the BB84 fixed-QBER analysis, the additive finite-size terms that are
independent of $q$ cancel in this difference, so that
$\Delta r_6(\alpha)$ directly isolates the benefit associated with trusted
local randomization.}

\subsection{Results}

\begin{figure}[t]
    \centering
    \includegraphics[width=\linewidth]{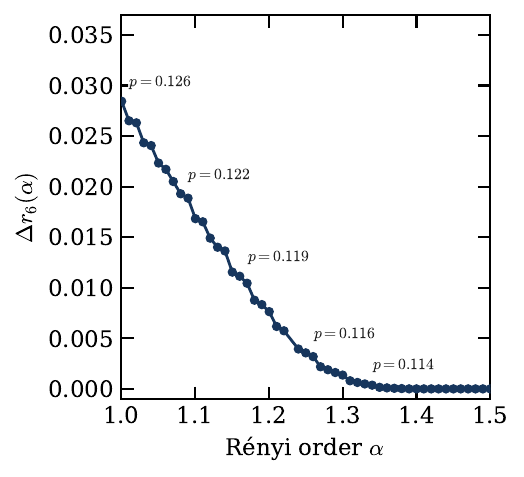}
    \caption{\justifying
    {Largest difference between the secret-key rate obtained with
    optimized trusted local randomization and that obtained for $q=0$ in
    the six-state protocol, plotted versus the R\'enyi order $\alpha$.
    The QBER is assumed to be known exactly and satisfies
    $Q_{\mathbb{X}}=Q_{\mathbb{Y}}=Q_{\mathbb{Z}}=p$. For the security budget $\epsilon_{\rm tot}=10^{-10}$, we set $\epsilon_{\rm PE}=\epsilon_{\rm PA}=\epsilon_{\rm EC}=\epsilon_{\rm tot}/3$; the parameter-estimation penalty is omitted under the exact-QBER assumption.}}
    \label{fig:sixstate_delta}
\end{figure}

{Figure~\ref{fig:sixstate_delta} shows that the dependence of the
trusted-noise advantage on the R\'enyi order closely resembles that found
for BB84. The largest enhancement is again obtained close to the
von Neumann limit, $\alpha\rightarrow1$. At the smallest value considered,
$\alpha=1.0005$, we find
$\Delta r_6\simeq2.84\times10^{-2}$, with the maximum attained at
$p\simeq0.126$. For BB84, the corresponding optimization gives
$\Delta r\simeq2.97\times10^{-2}$ at $p\simeq0.110$
[Fig.~\ref{fig:DeltaR}]. Thus, the maximum increase in secret-key rate
provided by local randomization is of essentially the same magnitude in
the two protocols, while in the six-state protocol it occurs at a larger
channel disturbance.}

{The latter behavior reflects the additional information supplied by the third measurement basis. Imposing the observed $\mathbb Y$-basis error rate further restricts the set of bipartite states, and hence of attacks, compatible with the measurement data. This reduces Eve's freedom in the worst-case optimization and allows the six-state protocol to remain secure at larger QBER. Trusted randomization consequently becomes most relevant in
a correspondingly higher-noise region. Nevertheless, the dependence on
the R\'enyi order remains qualitatively unchanged: as $\alpha$ increases,
the advantage decreases and eventually vanishes around
$\alpha\simeq1.4$, where the optimum returns to $q^*=0$. This is the same
transition observed for BB84 in Fig.~\ref{fig:DeltaR}.}

\begin{figure}[t]
    \centering
    \includegraphics[width=\linewidth]{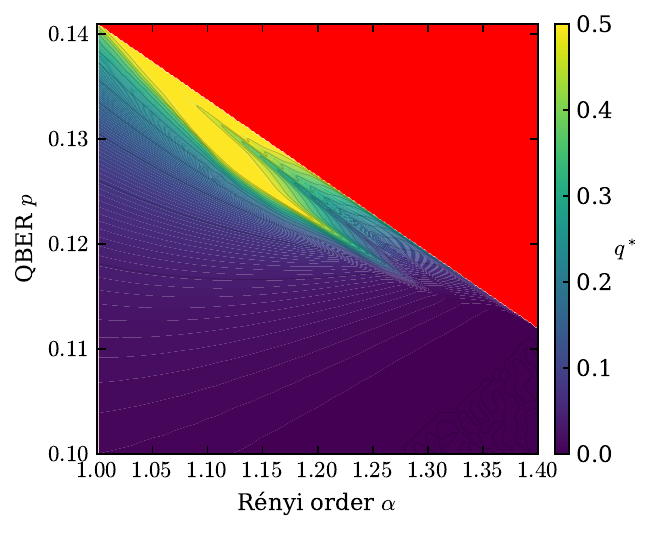}
    \caption{\justifying
    {Optimal trusted-noise probability $q^*(\alpha,p)$ for the
    six-state protocol under the exact-QBER condition
    $Q_{\mathbb{X}}=Q_{\mathbb{Y}}=Q_{\mathbb{Z}}=p$. The plot is restricted to the positive-key-rate
    region. As in BB84, trusted randomization becomes increasingly relevant
    as the disturbance approaches the boundary of this region. The same security budget $\epsilon_{\rm tot}=10^{-10}$ is divided as $\epsilon_{\rm PE}=\epsilon_{\rm PA}=\epsilon_{\rm EC}=\epsilon_{\rm tot}/3$; the parameter-estimation penalty is omitted.}}
    \label{fig:sixstate_qstar}
\end{figure}

{The optimal-noise map in Fig.~\ref{fig:sixstate_qstar} provides the
same physical picture. At small QBER the optimum remains close to
$q^*=0$, whereas the optimal flipping probability grows as the
disturbance approaches the largest values compatible with secret-key
generation. Compared with the BB84 map in Fig.~\ref{3D}, this structure is
shifted towards larger values of $p$, consistently with the greater noise
tolerance of the six-state protocol.}

{In particular, we recover limiting QBER values of approximately
$p\simeq0.126$ without preprocessing and $p\simeq0.141$ when the trusted
flipping probability is optimized. These values are consistent with the
enhanced noise tolerance of the six-state protocol previously obtained
for one-way key distillation with noisy preprocessing in
Ref.~\cite{renner2005information}. For comparison, the corresponding
values for BB84 are approximately $0.11$ and $0.124$, respectively.}

{The six-state extension therefore confirms that the main mechanism
identified for BB84 is not protocol-specific. Trusted local randomization
has essentially no effect at low disturbance, becomes advantageous close
to the security boundary of the protocol, and provides its largest gain
for R\'enyi orders close to $\alpha=1$. At the same time, the disappearance
of the advantage around $\alpha\simeq1.4$ persists after the additional
$\mathbb Y$-basis constraint is imposed.}

\section{{Asymmetric bit and phase errors}}
\label{app:asymmetric_qber}

{The analysis in the main text assumes a symmetric BB84 channel,
$\hat Q_{\mathbb{X}}=\hat Q_{\mathbb{Z}}$. To investigate whether the effect of trusted
preprocessing depends on the physical origin of the observed disturbance,
we now relax this assumption and consider independently observed error rates
$\hat Q_{\mathbb{X}}$ and $\hat Q_{\mathbb{Z}}$. Since the raw key is generated from the
$\mathbb Z$ basis, $Q_{\mathbb{Z}}$ represents the bit-error rate of the raw key,
whereas $Q_{\mathbb{X}}$ plays the role of the complementary, or phase-error,
constraint entering the privacy analysis~\cite{shor2000simple}.}

{We retain the same finite-sample parameter-estimation model used in
Sec.~\ref{sec:results}, but construct independent confidence intervals for
the two QBERs. With the parameter-estimation sample equally divided between
the two bases, the statistical deviation is}
{
\begin{align}
    \delta(N,f)
    =
    \sqrt{
        \frac{\ln(4/\epsilon_{\rm PE})}{fN}
    },
\end{align}
}
{and, for $b\in\{\mathbb X,\mathbb Z\}$,}
{
\begin{align}
    Q_b^-
    &=
    \max\!\left\{
        0,\hat Q_b-\delta(N,f)
    \right\},
    \nonumber\\
    Q_b^+
    &=
    \min\!\left\{
        \frac12,\hat Q_b+\delta(N,f)
    \right\}.
    \label{eq:asym_intervals}
\end{align}
}
{Hence, unlike the symmetric analysis, where the optimization is
restricted to the diagonal $Q_{\mathbb{X}}=Q_{\mathbb{Z}}$, the state is now allowed to vary
inside the rectangular confidence region
$[Q_{\mathbb{X}}^-,Q_{\mathbb{X}}^+]\times[Q_{\mathbb{Z}}^-,Q_{\mathbb{Z}}^+]$.}

{For fixed $(N,f,q,\alpha,\hat Q_{\mathbb{X}},\hat Q_{\mathbb{Z}})$, the corresponding
R\'enyiQKD conic problem is}
{
\begin{align}
    h^\star_{\beta,\mathrm{asym}}
    (q;N,f,\hat Q_{\mathbb{X}},\hat Q_{\mathbb{Z}})
    ={}&
    \min_{h,\rho,\sigma}\quad h
    \nonumber\\[-1mm]
    \mathrm{s.t.}\quad
    &\operatorname{tr}_B\rho
    =
    \frac{\mathbb I_A}{2},
    \qquad
    \operatorname{tr}\sigma=1,
    \nonumber\\
    &Q_{\mathbb{X}}^-
    \le
    Q_{\mathbb{X}}(\rho)
    \le
    Q_{\mathbb{X}}^+,
    \nonumber\\
    &Q_{\mathbb{Z}}^-
    \le
    Q_{\mathbb{Z}}(\rho)
    \le
    Q_{\mathbb{Z}}^+,
    \nonumber\\
    &(h,\rho,\sigma)
    \in
    \mathcal K_{\mathrm{R\acute enyiQKD}}
    ^{\beta,\mathrm{id},\mathcal Z,K}.
    \label{eq:conic_problem_asymmetric}
\end{align}
}
{The privacy contribution is recovered, as in
Eq.~\eqref{eq:conic_to_entropy}, according to}
{
\begin{align}
    d_{\beta,\mathrm{asym}}
    =
    \frac{
        \log_2[-h^\star_{\beta,\mathrm{asym}}]
    }{\beta-1}.
    \label{eq:dbeta_asymmetric}
\end{align}
}

{The error-correction term, however, depends directly on the
$\mathbb Z$-basis error. Since Bob flips every raw-key bit independently
with probability $q$, the worst-case bit-error probability after trusted
preprocessing is}
{
\begin{align}
    s_{\mathbb Z}(q)
    =
    q+(1-2q)Q_{\mathbb{Z}}^+ .
    \label{eq:asym_effective_bit_error}
\end{align}
}
{The corresponding finite-size rate is therefore}
{
\begin{align}
    r_{\mathrm{asym}}
    &=
    (1-f)
    \left[
        d_{\beta,\mathrm{asym}}
        -h_2\!\left(s_{\mathbb Z}(q)\right)
    \right]
    -
    \frac{
        g_{\epsilon_{\rm PA}}(\alpha)+t_{\rm EC}
    }{N}.
    \label{eq:rate_asymmetric}
\end{align}
}
{For every $N$ and pair $(\hat Q_{\mathbb{X}},\hat Q_{\mathbb{Z}})$ we optimize
Eq.~\eqref{eq:rate_asymmetric} over $f$, $\alpha$, and $q$. The reference
branch without trusted preprocessing is obtained by fixing $q=0$ and
optimizing independently over $f$ and $\alpha$.}

\subsection{Results}

\begin{figure}[t]
    \centering
    \includegraphics[width=\linewidth]{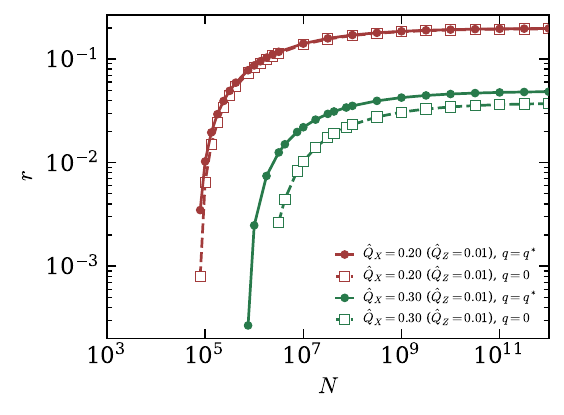}
    \caption{\justifying
    {Finite-size secret-key rate for asymmetric BB84 errors with
    $\hat Q_{\mathbb Z}=0.01$ and $\hat Q_{\mathbb X}=0.20,0.30$.
    Solid curves correspond to the optimized trusted-noise probability
    $q^*$, whereas dashed curves are obtained for $q=0$. Parameter
    estimation is included; $f$ and $\alpha$ are optimized independently
    for each branch. The security budget is $\epsilon_{\rm tot}=10^{-10}$,
    with $\epsilon_{\rm PE}=\epsilon_{\rm PA}=\epsilon_{\rm EC}=\epsilon_{\rm tot}/3$.}}
    \label{fig:asym_qx}
\end{figure}

{Figure~\ref{fig:asym_qx} illustrates the favorable case with small
bit error and large phase error. At $\hat Q_{\mathbb X}=0.20$, trusted
randomization improves the rate near the positive-key threshold, with a
gain that decreases at large $N$. At $\hat Q_{\mathbb X}=0.30$, the
improvement is more marked: the minimum block size falls from
$N\simeq3.2\times10^6$ at $q=0$ to $N\simeq7.5\times10^5$ with
optimized preprocessing. At $N=10^8$, the respective rates are
$2.33\times10^{-2}$ and $3.53\times10^{-2}$.}

\begin{figure}[t]
    \centering
    \includegraphics[width=\linewidth]{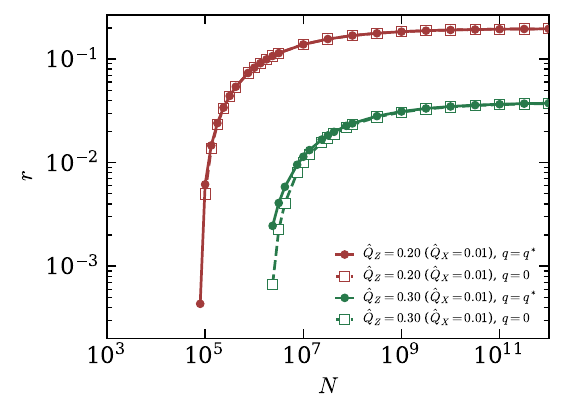}
    \caption{\justifying
    {Finite-size secret-key rate for the complementary asymmetric
    configuration, with $\hat Q_{\mathbb X}=0.01$ and
    $\hat Q_{\mathbb Z}=0.20,0.30$. The notation and optimization are
    those of Fig.~\ref{fig:asym_qx}. The security budget is
    $\epsilon_{\rm tot}=10^{-10}$, with
    $\epsilon_{\rm PE}=\epsilon_{\rm PA}=\epsilon_{\rm EC}=\epsilon_{\rm tot}/3$.}}
    \label{fig:asym_qz}
\end{figure}

{The reverse case, $\hat Q_{\mathbb X}=0.01$, is shown in
Fig.~\ref{fig:asym_qz}. For $\hat Q_{\mathbb Z}=0.20$, the optimized
branch becomes positive around $N\simeq7.9\times10^4$ rather than
$10^5$, but the advantage soon disappears and $q^*=0$ at larger $N$.
For $\hat Q_{\mathbb Z}=0.30$, both branches become positive around
$N\simeq2.4\times10^6$; at $N=10^8$ their rates are only
$2.39\times10^{-2}$ and $2.32\times10^{-2}$, respectively. }

{The different roles of the two errors in Eq.~\eqref{eq:rate_asymmetric}
explain why trusted preprocessing need not have the same effect when they
are interchanged. The familiar asymptotic BB84 rate without local
randomization, $r_\infty\ge 1-h_2(Q_{\mathbb X})-h_2(Q_{\mathbb Z})$
\cite{shor2000simple,kraus2005lower}, makes their interpretation
transparent: the $\mathbb X$-basis error acts as a phase-error constraint
on the privacy contribution $H(Y|E)$, whereas the $\mathbb Z$-basis error
sets the cost of reconciling the raw key. In particular, a small
$Q_{\mathbb X}$ already guarantees high conditional entropy of the
$\mathbb Z$-basis key, leaving little Eve information for trusted
randomization to suppress. A large $Q_{\mathbb X}$ instead permits a
larger privacy penalty, making a gain from randomization possible. This
separation is only interpretative: both observed errors enter the conic
minimization of $d_{\beta,\mathrm{asym}}$, while $Q_{\mathbb Z}$ also
appears explicitly in the reconciliation term.}

{Indeed, for $Q_{\mathbb Z}<1/2$ the flipped-bit error
$s_{\mathbb Z}(q)=q+(1-2q)Q_{\mathbb Z}^+$ increases towards $1/2$ as $q$ grows.
Thus, when $Q_{\mathbb X}$ is large but $Q_{\mathbb Z}$ is small, the
possible improvement in $H(Y|E)$ can compensate for the moderate
increase in $h_2(s_{\mathbb Z})$. In the opposite configuration, the phase-error
constraint already limits Eve's information, while a large bit-error
rate makes reconciliation costly; consequently the optimum approaches
$q=0$. Strongly asymmetric errors are physically possible: predominantly
Pauli-$Z$ (phase-flip) noise increases the error observed in the
$\mathbb X$ basis without comparably affecting the $\mathbb Z$ basis,
whereas predominantly Pauli-$X$ (bit-flip) noise produces the reverse.}

{Comparing the two figures shows that the advantage of trusted
preprocessing is controlled not just by the overall disturbance but by
its distribution between phase and bit errors.}

\section{Effect of trusted noise on min--entropy ($\alpha\to\infty$) secret-key rate}
\label{trust_min_ent}
Conditional min-entropy can be obtain from conditional sandwiched Rényi entropy taking the limit $\alpha\to\infty$ \cite{tomamichel2012framework}:
\begin{align}
    \tilde H^\uparrow_{\mathrm{min}}(A|B) &= \lim_{\alpha\to\infty} \tilde H^\uparrow_\alpha(A|B)\\
    &= 
    \sup_{\sigma_B}
    -\log\left\|
    \sigma_B^{-\frac{1}{2}}
    \rho_{AB}\,
    \sigma_B^{-\frac{1}{2}}
    \right \|_{\infty}
\end{align}
This entropic quantity has the fundamental operational meaning of a \emph{guessing probability} \cite{coles2012unification}
\begin{align}
     \tilde H^\uparrow_{\mathrm{min}}(Y|E) = -\log P_g^{(0)} (Y|E)
\end{align}
being $P_g^{(0)} (Y|E)$ the probability for an eavesdropper to guess the correct value of the random variable $Y$, having access to some side information available in the quantum system $E$. Now it is worth noting that the guessing probability has a simple analytical expression for BB84 protocol with symmetric QBER \cite{bratzik2011min,bunandar2020numerical,staffieri2026finite}
\begin{align}
    P_g^{(0)}(Y|E) = \frac12+\sqrt{p(1-p)}
\end{align}
with $p=Q_{\mathbb Z}=Q_{\mathbb X}$. Through leftover-hash lemma \cite{tomamichel2011leftover} the asymptotic secret-key rate under collective attacks can be bound by \cite{staffieri2026finite}:
\begin{align}
    r_{\mathrm{min}}&\le \tilde H^\uparrow_{\mathrm{min}}(Y|E)-\gamma h_2(p)\\
    &=-\log {P_g^{(0)}(Y|E)}-\gamma h_2(p)
\end{align}
In this particular case, we consider error correction efficiency $\gamma>1$, to be exhaustive. Now we observe that trusted noise alters Eve's guessing probability through a multiplicative factor $(1-2q)$, being $q$ Bob's flipping probability
\begin{align}
    P_g^{(q)}(Y|E) = \frac12 +(1-2q)\sqrt{p(1-p)}
\end{align}
Hence, if trusted local randomization has to be taken into account, the bound on the secret-key rate turns into
\begin{align}
     r_{\mathrm{min}}(q)&\le 
    -\log {P_g^{(q)}(Y|E)}-
    \gamma h_2\bigl(s(p,q)\bigr)
\end{align}
where $s(p,q)=p+q-2pq$ is the \emph{observed} QBER. Here we want to show that, for the secret-key rate estimator considered, adding trusted noise does not provide an advantage; mathematically this is equivalent to show that when $r_{\mathrm{min}}>0$, it is monotonically decreasing in $q\in[0,\tfrac{1}{2}]$.
\paragraph{Claim.} On the physically relevant domain (\emph{i.e.}, where the guessing probability bound is defined and $s(p,q)\le \tfrac12$), \textit{whenever $r_{\mathrm{min}}(q)>0$} one has
\begin{equation}
r'_{\mathrm{min}}(q) \le0.
\end{equation}
mathematically this statement translates into
\begin{align}
    \forall q\in[0,\tfrac12]
    \;\;\textrm{s.t.}\;\; 
    r_{\mathrm{min}}(q) \ge0\;: \qquad r'_{\mathrm{min}}(q) \le0
\end{align}
\paragraph{Proof.}
First of all we set $t=\sqrt{p(1-p)}$ and  $A(q)=\tfrac12+(1-2q)t$; then the first and second derivative of  $ r_{\mathrm{min}}(q)$ read
\begin{align}
     r'_{\mathrm{min}}(q) &= \frac{2t}{A(q)\,\ln 2}-
     \gamma(1-2p)\,\log\frac{1-s(q)}{s(q)}\\
     r''_{\mathrm{min}}(q) &=\frac{4t^2}{A(q)^2\,\ln 2}+
     \frac{\gamma(1-2p)^2}{\ln 2}
     \frac{1}{s(q)\,[1-s(q)]}
\end{align}
It is easy to see that $\forall p,q\in[0,\tfrac{1}{2}]:  r''_{\mathrm{min}}(q)>0$, then $ r_{\mathrm{min}}(q)$ is strictly convex in $q$ everywhere in the physical domain. Moreover at the upper boundary:
\begin{align}
r_{\mathrm{min}}\left(\tfrac12\right) 
&= 1-\gamma \;\le\; 0, 
\nonumber\\
r'_{\mathrm{min}}\left(\tfrac12\right) &=
\frac{4\sqrt{p(1-p)}}{\ln 2} > 0 
\quad (0<p<\tfrac12).
\end{align}
Since $r_{\mathrm{min}}$ is convex and $r_{\mathrm{min}}(\tfrac12)\le 0$ with $r_{\mathrm{min}}'(\tfrac12)>0$, any region where $r_{\mathrm{min}}(q)>0$ must lie to the left of the (at most single) zero of $r_{\mathrm{min}}'$. Let us indicate $q_0$ the unique root of $r_{\mathrm{min}}'$ if it exists. The point $q=q_0$ is a minimum for $r_{\mathrm{min}}$, then in $[q_0,\tfrac12]$   $r_{\mathrm{min}}$ is increasing and bounded above by $r_{\mathrm{min}}(\tfrac12)\le 0$, hence $r_{\mathrm{min}}\le 0$ there; therefore all points with $r_{\mathrm{min}}>0$ lie in $[0,q_0)$ where $r_{\mathrm{min}}'\le 0$. If $r_{\mathrm{min}}'$ had no zero and were nonnegative throughout, $r_{\mathrm{min}}$ would be increasing on $[0,\tfrac12]$ and thus $r_{\mathrm{min}}(\tfrac12)\ge r_{\mathrm{min}}(0)$, contradicting $r_{\mathrm{min}}(\tfrac12)\le 0$ when $r_{\mathrm{min}}(0)>0$. Hence, in every point with $r_{\mathrm{min}}(q)>0$ one necessarily has $r_{\mathrm{min}}'(q)\le 0$. \qed

\bibliography{bibliography}

@Article{Gisin2002,
  author    = {Gisin, Nicolas and Ribordy, Grégoire and Tittel, Wolfgang and Zbinden, Hugo},
  journal   = {Reviews of Modern Physics},
  title     = {Quantum cryptography},
  year      = {2002},
  issn      = {1539-0756},
  month     = mar,
  number    = {1},
  pages     = {145--195},
  volume    = {74},
  doi       = {10.1103/revmodphys.74.145},
  publisher = {American Physical Society (APS)},
}

@Article{Pirandola2020,
 author    = {Pirandola, S. and Andersen, U. L. and Banchi, L. and Berta, M. and Bunandar, D. and Colbeck, R. and Englund, D. and Gehring, T. and Lupo, C. and Ottaviani, C. and Pereira, J. L. and Razavi, M. and Shamsul Shaari, J. and Tomamichel, M. and Usenko, V. C. and Vallone, G. and Villoresi, P. and Wallden, P.},
  journal   = {Advances in Optics and Photonics},
  title     = {Advances in quantum cryptography},
  year      = {2020},
  issn      = {1943-8206},
  month     = dec,
  number    = {4},
  pages     = {1012},
  volume    = {12},
  doi       = {10.1364/aop.361502},
  publisher = {Optica Publishing Group},
}

@Article{Portmann2022,
  author    = {Portmann, Christopher and Renner, Renato},
  journal   = {Reviews of Modern Physics},
  title     = {Security in quantum cryptography},
  year      = {2022},
  issn      = {1539-0756},
  month     = jun,
  number    = {2},
  pages     = {025008},
  volume    = {94},
  doi       = {10.1103/revmodphys.94.025008},
  publisher = {American Physical Society (APS)},
}

@Article{Primaatmaja2023,
  author    = {Primaatmaja, Ignatius W. and Goh, Koon Tong and Tan, Ernest Y.-Z. and Khoo, John T.-F. and Ghorai, Shouvik and Lim, Charles C.-W.},
  journal   = {Quantum},
  title     = {Security of device-independent quantum key distribution protocols: a review},
  year      = {2023},
  issn      = {2521-327X},
  month     = mar,
  pages     = {932},
  volume    = {7},
  doi       = {10.22331/q-2023-03-02-932},
  publisher = {Verein zur Forderung des Open Access Publizierens in den Quantenwissenschaften},
}

@book{tomamichel2015quantum,
  title={Quantum information processing with finite resources: mathematical foundations},
  author={Tomamichel, Marco},
  volume={5},
  year={2015},
  publisher={Springer}
}

@article{renner2008security,
  title={Security of quantum key distribution},
  author={Renner, Renato},
  journal={International Journal of Quantum Information},
  volume={6},
  number={01},
  pages={1--127},
  year={2008},
  publisher={World Scientific}
}

@article{bennett2014quantum,
  title={Quantum cryptography: Public key distribution and coin tossing},
  author={Bennett, Charles H and Brassard, Gilles},
  journal={Theoretical computer science},
  volume={560},
  pages={7--11},
  year={2014},
  publisher={Elsevier}
}

@article{shor2000simple,
  title={Simple proof of security of the BB84 quantum key distribution protocol},
  author={Shor, Peter W and Preskill, John},
  journal={Physical review letters},
  volume={85},
  number={2},
  pages={441},
  year={2000},
  publisher={APS}
  }

@article{renner2005information,
  title={Information-theoretic security proof for quantum-key-distribution protocols},
  author={Renner, Renato and Gisin, Nicolas and Kraus, Barbara},
  journal={Physical Review A—Atomic, Molecular, and Optical Physics},
  volume={72},
  number={1},
  pages={012332},
  year={2005},
  publisher={APS}
}

@article{kraus2005lower,
  title={Lower and Upper Bounds on the Secret-Key Rate for Quantum Key Distribution Protocols Using One-Way Classical Communication},
  author={Kraus, Barbara and Gisin, Nicolas and Renner, Renato},
  journal={Physical review letters},
  volume={95},
  number={8},
  pages={080501},
  year={2005},
  publisher={APS}
}

@article{kamin2025r,
  title={R$\backslash$'enyi security framework against coherent attacks applied to decoy-state QKD},
  author={Kamin, Lars and Burniston, John and Tan, Ernest Y-Z},
  journal={arXiv preprint arXiv:2504.12248},
  year={2025}
}

@article{Kamin_2025,
   title={Finite-Size Analysis of Prepare-and-Measure and Decoy-State Quantum Key Distribution via Entropy Accumulation},
   volume={6},
   ISSN={2691-3399},
   url={http://dx.doi.org/10.1103/PRXQuantum.6.020342},
   DOI={10.1103/prxquantum.6.020342},
   number={2},
   journal={PRX Quantum},
   publisher={American Physical Society (APS)},
   author={Kamin, Lars and Arqand, Amir and George, Ian and Lütkenhaus, Norbert and Tan, Ernest Y.-Z.},
   year={2025},
   month=jun }

@article{staffieri2026finiteCV,
  title={Finite-size secret-key rates of discrete modulation continuous-variable quantum key distribution under Gaussian attacks},
  author={Staffieri, Gabriele and Scala, Giovanni and Lupo, Cosmo},
  journal={Physical Review A},
  volume={113},
  number={2},
  pages={022445},
  year={2026},
  publisher={APS}
}

@article{dupuis2023privacy,
  title={Privacy amplification and decoupling without smoothing},
  author={Dupuis, Fr{\'e}d{\'e}ric},
  journal={IEEE Transactions on Information Theory},
  volume={69},
  number={12},
  pages={7784--7792},
  year={2023},
  publisher={IEEE}
}

@article{george2025finite,
  title={Finite-key analysis of quantum key distribution with characterized devices using entropy accumulation},
  author={George, Ian and Lin, Jie and van Himbeeck, Thomas and Fang, Kun and L{\"u}tkenhaus, Norbert},
  journal={Quantum},
  volume={9},
  pages={1941},
  year={2025},
  publisher={Verein zur F{\"o}rderung des Open Access Publizierens in den Quantenwissenschaften}
}

@article{winick2018reliable,
  title={Reliable numerical key rates for quantum key distribution},
  author={Winick, Adam and L{\"u}tkenhaus, Norbert and Coles, Patrick J},
  journal={Quantum},
  volume={2},
  pages={77},
  year={2018},
  publisher={Verein zur F{\"o}rderung des Open Access Publizierens in den Quantenwissenschaften}
}

@article{tomamichel2011leftover,
  title={Leftover hashing against quantum side information},
  author={Tomamichel, Marco and Schaffner, Christian and Smith, Adam and Renner, Renato},
  journal={IEEE Transactions on Information Theory},
  volume={57},
  number={8},
  pages={5524--5535},
  year={2011},
  publisher={IEEE}
}

@article{tomamichel2014relating,
  title={Relating different quantum generalizations of the conditional R{\'e}nyi entropy},
  author={Tomamichel, Marco and Berta, Mario and Hayashi, Masahito},
  journal={Journal of Mathematical Physics},
  volume={55},
  number={8},
  year={2014},
  publisher={AIP Publishing}
}

@article{chung2025generalized,
  title={Generalized numerical framework for improved finite-sized key rates with R{\'e}nyi entropy},
  author={Chung, Rebecca RB and Ng, Nelly HY and Cai, Yu},
  journal={Physical Review A},
  volume={112},
  number={1},
  pages={012612},
  year={2025},
  publisher={APS}
}

@book{bhatia2013matrix,
  title={Matrix analysis},
  author={Bhatia, Rajendra},
  volume={169},
  year={2013},
  publisher={Springer Science \& Business Media}
}

@article{rubboli2022new,
  title={New additivity properties of the relative entropy of entanglement and its generalizations},
  author={Rubboli, Roberto and Tomamichel, Marco},
  journal={arXiv preprint arXiv:2211.12804},
  year={2022}
}

@article{tomamichel2012framework,
  title={A framework for non-asymptotic quantum information theory},
  author={Tomamichel, Marco},
  journal={arXiv preprint arXiv:1203.2142},
  year={2012}
}

@article{staffieri2026finite,
  title={Finite-size security of QKD: comparison of three proof techniques},
  author={Staffieri, Gabriele and Scala, Giovanni and Lupo, Cosmo},
  journal={The European Physical Journal Plus},
  volume={141},
  number={6},
  pages={754},
  year={2026},
  publisher={Springer}
}

@article{coles2012unification,
  title={Unification of different views of decoherence and discord},
  author={Coles, Patrick J},
  journal={Physical Review A—Atomic, Molecular, and Optical Physics},
  volume={85},
  number={4},
  pages={042103},
  year={2012},
  publisher={APS}
}

@article{bratzik2011min,
  title={Min-entropy and quantum key distribution: Nonzero key rates for “small” numbers of signals},
  author={Bratzik, Sylvia and Mertz, Markus and Kampermann, Hermann and Bru{\ss}, Dagmar},
  journal={Physical Review A—Atomic, Molecular, and Optical Physics},
  volume={83},
  number={2},
  pages={022330},
  year={2011},
  publisher={APS}
}

@article{bunandar2020numerical,
  title={Numerical finite-key analysis of quantum key distribution},
  author={Bunandar, Darius and Govia, Luke CG and Krovi, Hari and Englund, Dirk},
  journal={npj Quantum Information},
  volume={6},
  number={1},
  pages={104},
  year={2020},
  publisher={Nature Publishing Group UK London}
}

@article{renes2007noisy,
  title={Noisy processing and distillation of private quantum states},
  author={Renes, Joseph M and Smith, Graeme},
  journal={Physical review letters},
  volume={98},
  number={2},
  pages={020502},
  year={2007},
  publisher={APS}
}

@article{mertz2013quantum,
  title={Quantum key distribution with finite resources: Taking advantage of quantum noise},
  author={Mertz, Markus and Kampermann, Hermann and Shadman, Zahra and Bru{\ss}, Dagmar},
  journal={Physical Review A—Atomic, Molecular, and Optical Physics},
  volume={87},
  number={4},
  pages={042312},
  year={2013},
  publisher={APS}
}

@article{navarro2025finite,
  title   = {Finite-Size Quantum Key Distribution Rates from {R\'enyi} Entropies Using Conic Optimization},
  author  = {Navarro, Mariana and Gonz{\'a}lez Lorente, Andr{\'e}s and Parellada, Pablo V. and Pascual-Garc{\'i}a, Carlos and Ara{\'u}jo, Mateus},
  journal = {PRX Quantum},
  volume  = {7},
  pages   = {033051},
  year    = {2026},
  doi     = {10.1103/bf9s-m4jb},
  url     = {https://doi.org/10.1103/bf9s-m4jb}
}

@article{lorente2025quantum,
  title={Quantum key distribution rates from non-symmetric conic optimization},
  author={Lorente, Andr{\'e}s Gonz{\'a}lez and Parellada, Pablo V and Castillo-Celeita, Miguel and Ara{\'u}jo, Mateus},
  journal={Quantum},
  volume={9},
  pages={1657},
  year={2025},
  publisher={Verein zur F{\"o}rderung des Open Access Publizierens in den Quantenwissenschaften}
}

@article{zhang2024device,
  title={Device-independent quantum secret sharing with noise preprocessing and postselection},
  author={Zhang, Qi and Zhong, Wei and Du, Ming-Ming and Shen, Shu-Ting and Li, Xi-Yun and Zhang, An-Lei and Zhou, Lan and Sheng, Yu-Bo},
  journal={Physical Review A},
  volume={110},
  number={4},
  pages={042403},
  year={2024},
  publisher={APS}
}

@Article{Ghoreishi2025a,
  author    = {Ghoreishi, Seyed Arash and Scala, Giovanni and Renner, Renato and Tacca, Letícia Lira and Bouda, Jan and Walborn, Stephen Patrick and Pawłowski, Marcin},
  journal   = {Physics Reports},
  title     = {The future of secure communications: Device independence in quantum key distribution},
  year      = {2025},
  issn      = {0370-1573},
  month     = dec,
  pages     = {1--97},
  volume    = {1149},
  doi       = {10.1016/j.physrep.2025.09.006},
  publisher = {Elsevier BV},
}

@article{tan2022improved,
  title={Improved DIQKD protocols with finite-size analysis},
  author={Tan, Ernest Y-Z and Sekatski, Pavel and Bancal, Jean-Daniel and Schwonnek, Ren{\'e} and Renner, Renato and Sangouard, Nicolas and Lim, Charles C-W},
  journal={Quantum},
  volume={6},
  pages={880},
  year={2022},
  publisher={Verein zur F{\"o}rderung des Open Access Publizierens in den Quantenwissenschaften}
}

@article{yamano2023general,
  title={General treatment of Gaussian trusted noise in continuous variable quantum key distribution},
  author={Yamano, Shinichiro and Matsuura, Takaya and Kuramochi, Yui and Sasaki, Toshihiko and Koashi, Masato},
  journal={arXiv preprint arXiv:2305.17684},
  year={2023}
}

@article{ulu2025device,
  title={Device Independent Quantum Key Activation},
  author={Ulu, Bora and Brunner, Nicolas and Weilenmann, Mirjam},
  journal={Physical Review Letters},
  volume={135},
  number={19},
  pages={190801},
  year={2025}
}

@article{regula2026rethinking,
  title={Rethinking quantum smooth entropies: Tight one-shot analysis of quantum privacy amplification},
  author={Regula, Bartosz and Tomamichel, Marco},
  journal={arXiv preprint arXiv:2603.04493},
  year={2026}
}

@misc{GG84,
  author       = {Staffieri, Gabriele and D'Ambruoso, Giuseppe and Scala, Giovanni and Lupo, Cosmo},
  title        = {Python code for analysis of finite-size BB84 protocol with trusted noise preprocessing},
  url = {https://github.com/giuseppedambruoso/trusted_noise_qkd},
  month        = mar,
  year         = 2026
}

@misc{ConicQKDJulia,
  author       = {Ara{\'u}jo, Mateus and {ConicQKD.jl contributors}},
  title        = {{ConicQKD.jl}: Implementation of convex cones for quantum key distribution},
  howpublished = {\url{https://github.com/araujoms/ConicQKD.jl}},
  year         = {2026},
  note         = {Julia software package}
}

@article{KernRenes2008,
  author  = {Kern, Oliver and Renes, Joseph M.},
  title   = {Improved one-way rates for {BB84} and 6-state protocols},
  journal = {Quantum Information and Computation},
  volume  = {8},
  number  = {8-9},
  pages   = {756--772},
  year    = {2008},
  doi     = {10.26421/QIC8.8-9-6}
}

@article{SmithRenesSmolin2008,
  author  = {Smith, Graeme and Renes, Joseph M. and Smolin, John A.},
  title   = {Structured Codes Improve the {Bennett-Brassard-84}
             Quantum Key Rate},
  journal = {Physical Review Letters},
  volume  = {100},
  number  = {17},
  pages   = {170502},
  year    = {2008},
  doi     = {10.1103/PhysRevLett.100.170502}
}

@article{GottesmanLo2003,
  author  = {Gottesman, Daniel and Lo, Hoi-Kwong},
  title   = {Proof of Security of Quantum Key Distribution
             with Two-Way Classical Communications},
  journal = {IEEE Transactions on Information Theory},
  volume  = {49},
  number  = {2},
  pages   = {457--475},
  year    = {2003},
  doi     = {10.1109/TIT.2002.807289}
}

@misc{Navarro2026DMQKD,
  author        = {Navarro, Mariana and Ara{\'u}jo, Mateus and Ac{\'i}n, Antonio and Pascual-Garc{\'i}a, Carlos},
  title         = {Effective discrete-modulated continuous variable {QKD} under general attacks using dimension reduction},
  year          = {2026},
  eprint        = {2606.20346},
  archivePrefix = {arXiv},
  primaryClass  = {quant-ph},
  doi           = {10.48550/arXiv.2606.20346},
  url           = {https://arxiv.org/abs/2606.20346}
}

@misc{TrustedNoiseQKD2026,
  author       = {Staffieri, Gabriele and D'Ambruoso, Giuseppe
                  and Scala, Giovanni and Lupo, Cosmo},
  title        = {{TrustedNoiseQKD.jl}: Finite-size quantum key
                  distribution with trusted noisy preprocessing},
  year         = {2026},
  howpublished = {GitHub repository},
  url          = {https://github.com/gabrielestaffieri/TrustedNoiseQKD.jl},
  note         = {Julia software, version 0.1.0}
}

@misc{PascualGarcia2026RPSK,
  author        = {Pascual-Garc{\'i}a, Carlos},
  title         = {Finite-size general security for relativistic
                   phase shift keying via variable-length
                   quantum key distribution},
  year          = {2026},
  eprint        = {2605.06249},
  archivePrefix = {arXiv},
  primaryClass  = {quant-ph},
  doi           = {10.48550/arXiv.2605.06249},
  url           = {https://arxiv.org/abs/2605.06249v2},
  note          = {arXiv:2605.06249v2}
}
\end{document}